\documentclass[useAMS,usenatbib]{aa} 
\usepackage{graphicx}
\usepackage{hyperref}
\usepackage{txfonts}
\usepackage{float}
\usepackage{commath}
\usepackage{wrapfig}
\usepackage{natbib}
\bibpunct{(}{)}{;}{a}{}{,} 

\newcommand{\kms}{$\rm km\,s^{-1}$}

\def\kms{$\rm km\;s^{-1}$}

\def\ha{H$\alpha$}
\def\hb{H$\beta$}

\def\oiii{[\ion{O}{iii}]}
\def\nii{[N~{\small II}]}

\def\mgb{Mg~{\it b}}
\def\fei{Fe{\small 5270}}
\def\feii{Fe{\small 5335}}
\def\feiii{Fe{\small 5406}}
\def\feiv{Fe{\small 5709}}
\def\fev{Fe{\small 5782}}
\newcommand{\TiOone}{\ensuremath{{\rm TiO}_1}}
\newcommand{\TiOtwo}{\ensuremath{{\rm TiO}_2}}
\newcommand{\Fe}{\ensuremath{\langle {\rm Fe}\rangle}}
\newcommand{\MgFep}{\ensuremath{[{\rm MgFe}]^{\prime}}}

\begin{document} 
   \title{The properties of the kinematically distinct components in NGC~448 and NGC~4365 \thanks{This work is based on observations taken at ESO La Silla Parnal Observatory within the program 094.B-0225(A)}
   }
     \author{
	 B. Nedelchev \inst{1,2}    
     \and L. Coccato \inst{2}
     \and E. M. Corsini \inst{3,4}
     \and M.\,~Sarzi\inst{1}
     \and T.\,~de~Zeeuw\inst{2,5,6}
     \and A. Pizzella \inst{3,4} 
     \and E. Dalla Bont\`a \inst{3,4} 
     \and E. Iodice \inst{7}
     \and L. Morelli \inst{8} 
     }    
  
  \institute{
    Centre for Astrophysics Research, University of Hertfordshire, College Lane, Hatfield AL10 9AB, UK\\
    \email{b.nedelchev@herts.ac.uk}
	\and ESO, Karl Schwarzschild Strasse 2, D-85748 Garching b. München, Germany 
    \and Dipartimento di Fisica e Astronomia ``G. Galilei'', Universit\`a di Padova, vicolo dell'Osservatorio 3, I-35122 Padova, Italy. 
    \and Max-Planck-Institut f\"ur extraterrestrische Physik, Giessenbachstrasse,  85741 Garching, Germany
    \and Sterrewacht Leiden, Leiden University, Postbus 9513, 2300 RA Leiden, The Netherlands
    \and INAF- Osservatorio Astronomico di Capodimonte, via Moiariello 16, I-80131 Napoli, Italy
    \and INAF-Osservatorio Astronomico di Padova, vicolo dell'Osservatorio 5, I-35122 Padova, Italy.
	\and Instituto de Astronomia y Ciencias Planetarias, Universidad de Atacama, Copiap\`o, Chile}
	\date{Received ...; accepted ...}

\newcommand{\IncludeTblone}{
\begin{table}
\caption{Photometric components in NGC~448 and NGC~4365. The first column gives the description of the fitted photometric component. Second column is the integrated magnitude of the photometric component returned by {\tt galfit} (see Sec. 3). The third column provides the estimated effective radius in arcseconds. The fourth column outlines the best-fitting S\`ersic index. The fifth column is the semi-minor to semi-major axial ratio of the best-fitting model. The last column is the position angle of the best-fitting S\`ersic profile.}   
\label{table:NGC448_PRS}      %
\centering                    %
\begin{tabular}{c c c c c c}      %
\hline\hline                  %
Name & mag & $R_{e}$ & n & b/a & PA \\ %
\hline                                 %
KDC (NGC~448)& 19.32 & 1.04 & 1.16 & 0.89 & -52.9 \\  %
KDC (NGC~448)& 17.25 & 6.50 & 1.03 & 0.28 & -61.0 \\
Main (NGC~448)& 16.11 & 22.40 & 1.89 & 0.89 & -65.0 \\
KDC (NGC~4365)& 6.70 & 27.72 & 1.46 & 0.77 & 44.3 \\  %
Main (NGC~4365)& 9.10 & 3.90 & 0.98 & 0.75 & 42.7\\
\hline                                 %
\end{tabular}
\end{table}
}

\newcommand{\IncludeFigOne}{
\begin{figure}
\centering 
\includegraphics[width=0.49\textwidth]{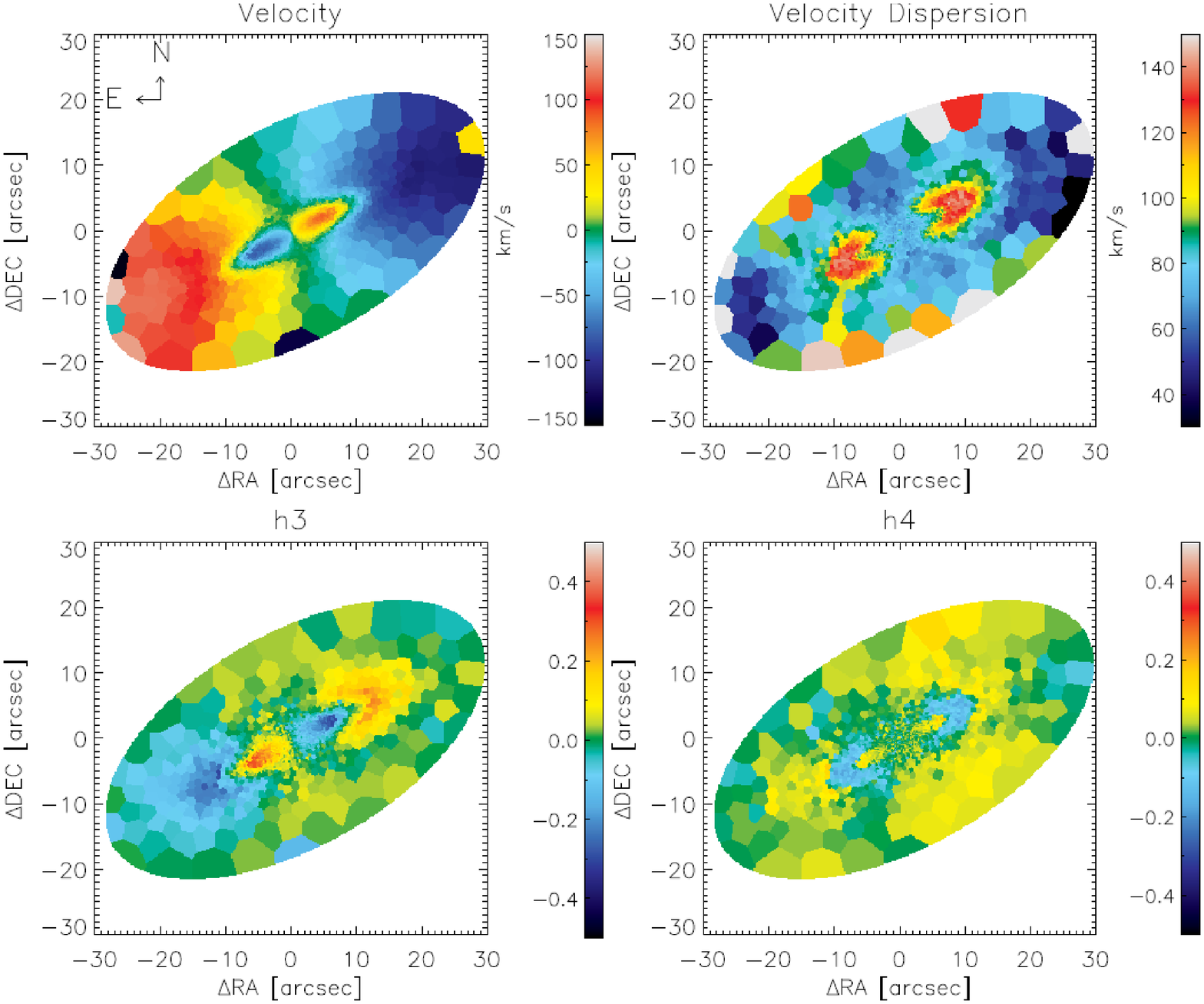}
\caption{The one-component kinematic maps for NGC~448. Top Left: The retrieved velocity field. Top Right: The velocity dispersion field. Bottom Left: The map of the {\tt h3} Gauss-Hermite coefficient in the parametrisation of the LOSVD. Bottom Right: The map of the {\tt h4} Gauss-Hermite coefficient in the parametrisation of the LOSVD.}
\label{fig:SPEC_NGC448} 
\end{figure} 
}  

\newcommand{\IncludeFigTwo}{
\begin{figure}
\centering 
\includegraphics[width=0.49\textwidth]{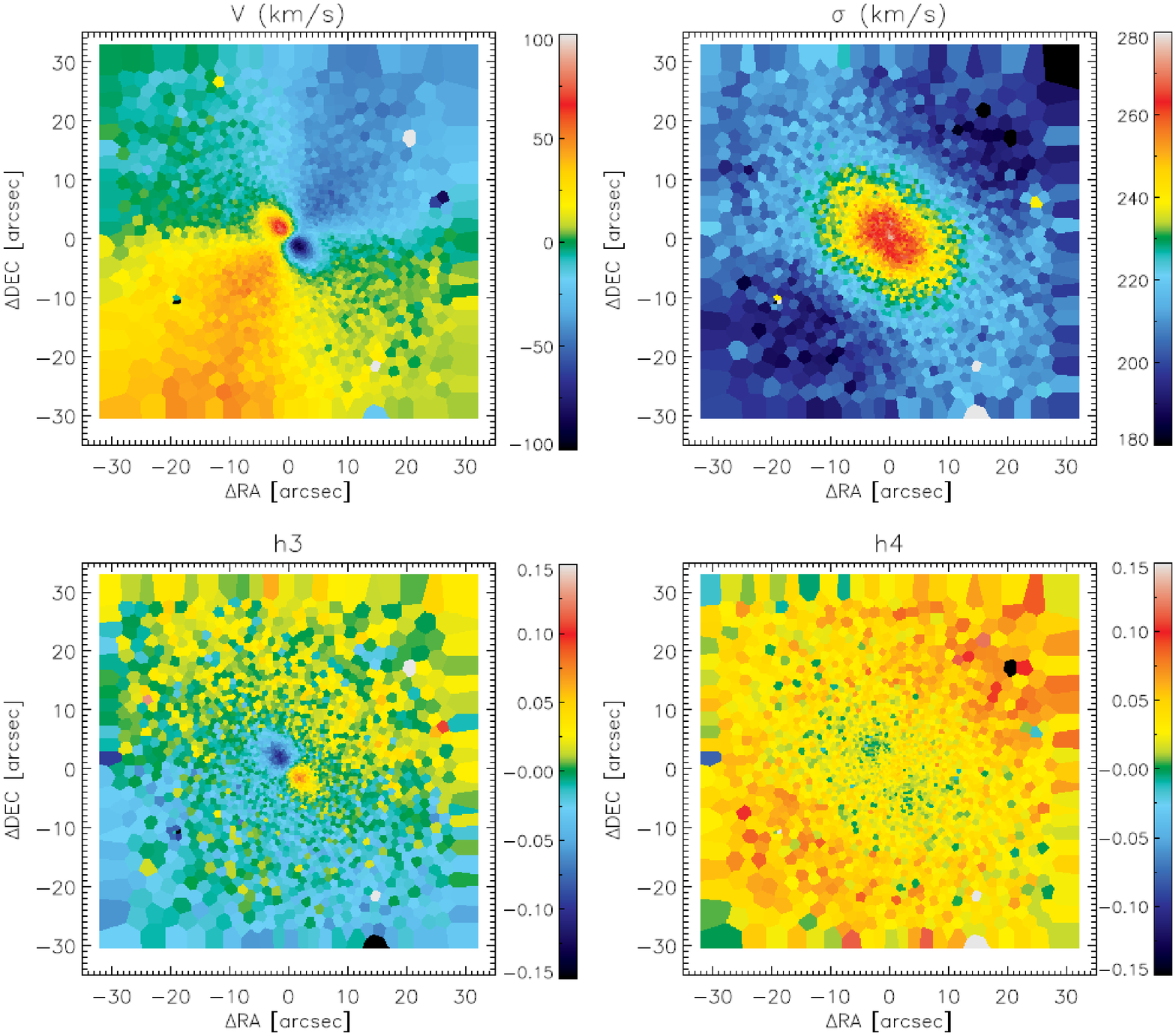}
\caption{The one-component kinematic maps for NGC~4365. Top Left: The retrieved velocity field. Top Right: The velocity dispersion field. Bottom Left: The map of the{\tt h3} Gauss-Hermite coefficient in the parametrisation of the LOSVD. Bottom Right: The map of the {\tt h4} Gauss-Hermite coefficient in the parametrisation of the LOSVD.}
\label{fig:SPEC_NGC4365} 
\end{figure} 
} 

\newcommand{\IncludeFigThree}{
\begin{figure}
\centering 
\includegraphics[width=0.46\textwidth]{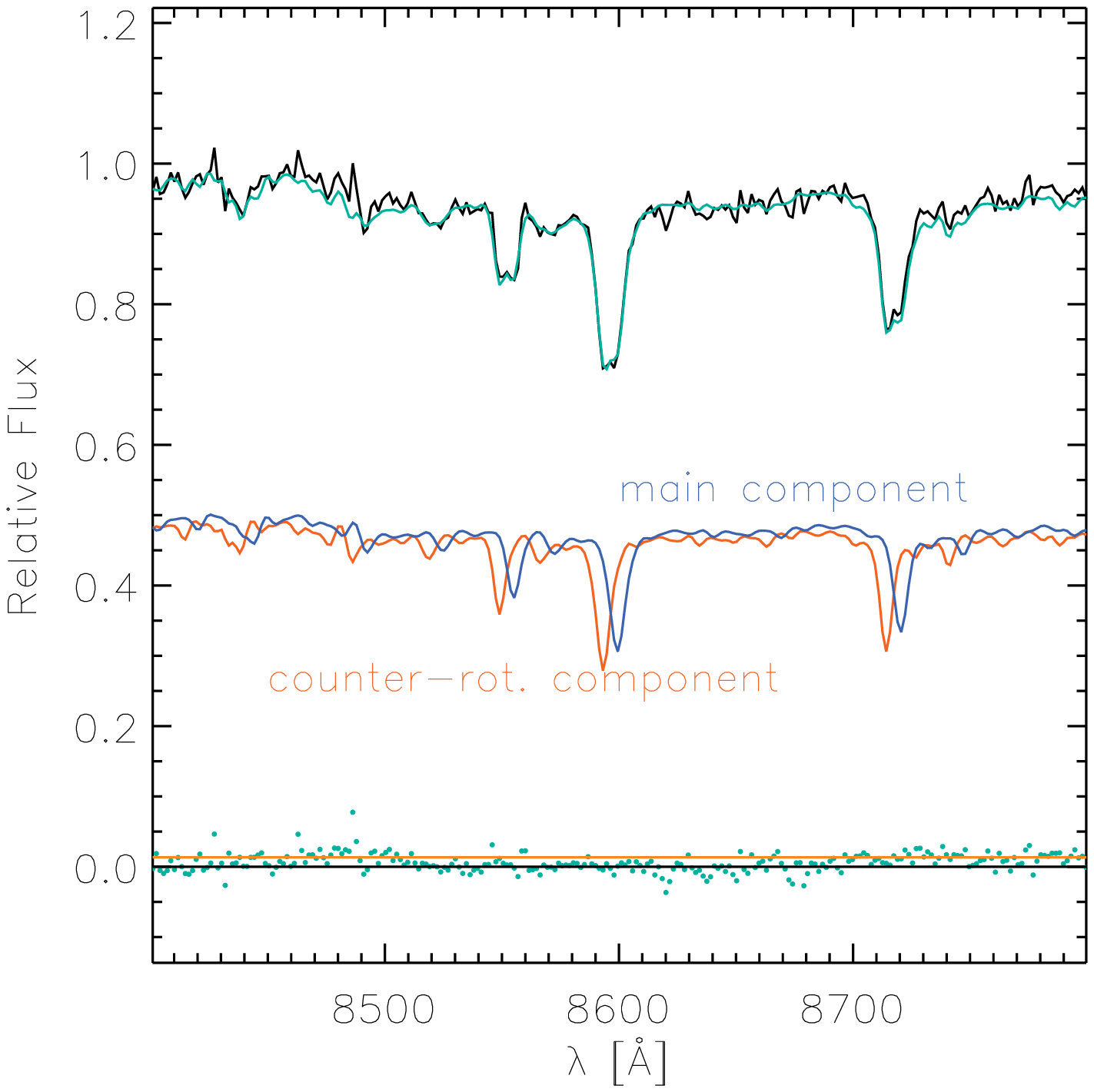}
\caption{Fit to the CaT spectral region for a bin approximately 9 arcsec roughly along the major axis of the centre of the NGC~448 field where the two distinct kinematic components have absolute velocity difference of $\sim$ 212\ \kms. The best-fitting model is in green. The contributions of the counter-rotating and main galaxy component are shown in blue and red respectively. The green points signify the residual from our fitting and the orange line is set at the level of the resistant dispersion of the residual.}
\label{fig:FIT_NGC448} 
\end{figure} 
}

\newcommand{\IncludeFigSix}{
\begin{figure*}
\centering 
\includegraphics[width=0.99\textwidth]{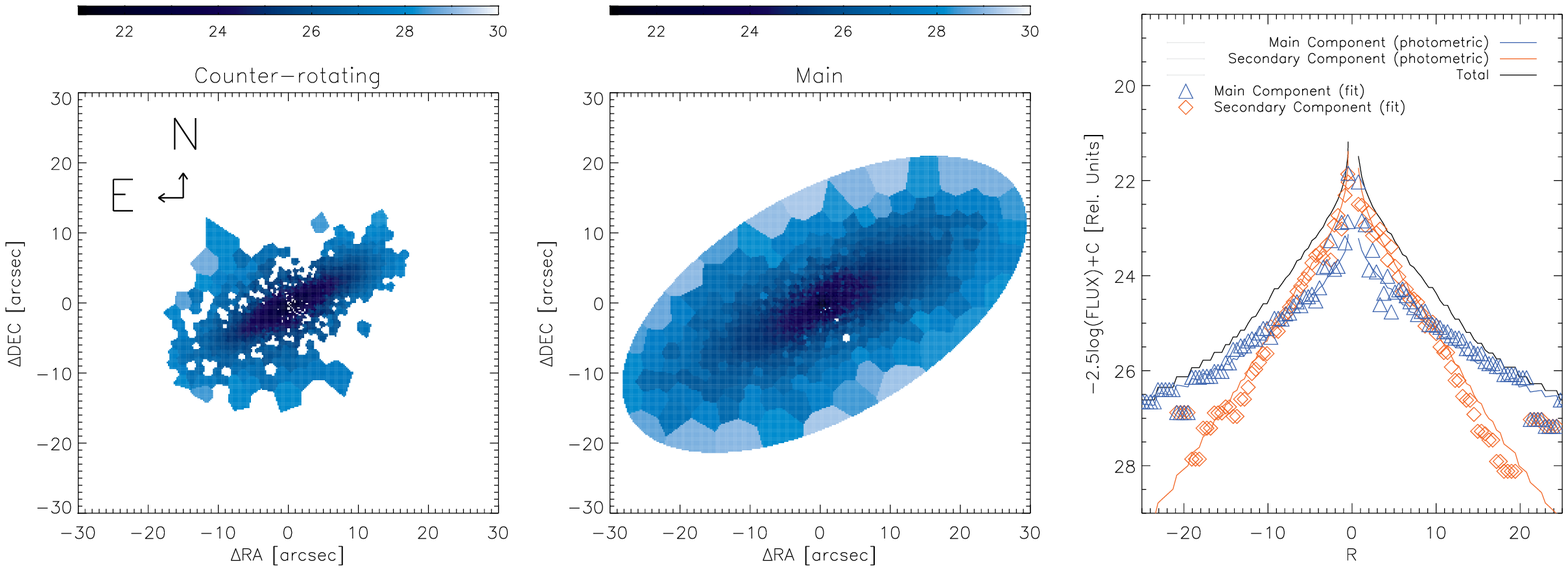}
\caption{Spectral flux maps and profile averaged over the number of spaxels for NGC 448. Left panel: Reconstructed surface brightness map of the counter-rotating component. Middle panel: Reconstructed surface brightness map of the main component. Right panel: Radial profile for both kinematic components, taken through a pseudo slit approximately aligned with the semi-major axis. The black, orange, and blue solid lines illustrate the photometrically derived total, counter-rotating, and main component flux contribution. The blue triangles and the orange diamonds show the flux contribution derived of the counter-rotating and main components as derived through our fitting procedure without imposing any constraints.}
\label{fig:IMG_FRAC_NGC448} 
\end{figure*} 
}  

\newcommand{\IncludeFigSeven}{
\begin{figure}
\centering 
\includegraphics[width=0.49\textwidth]{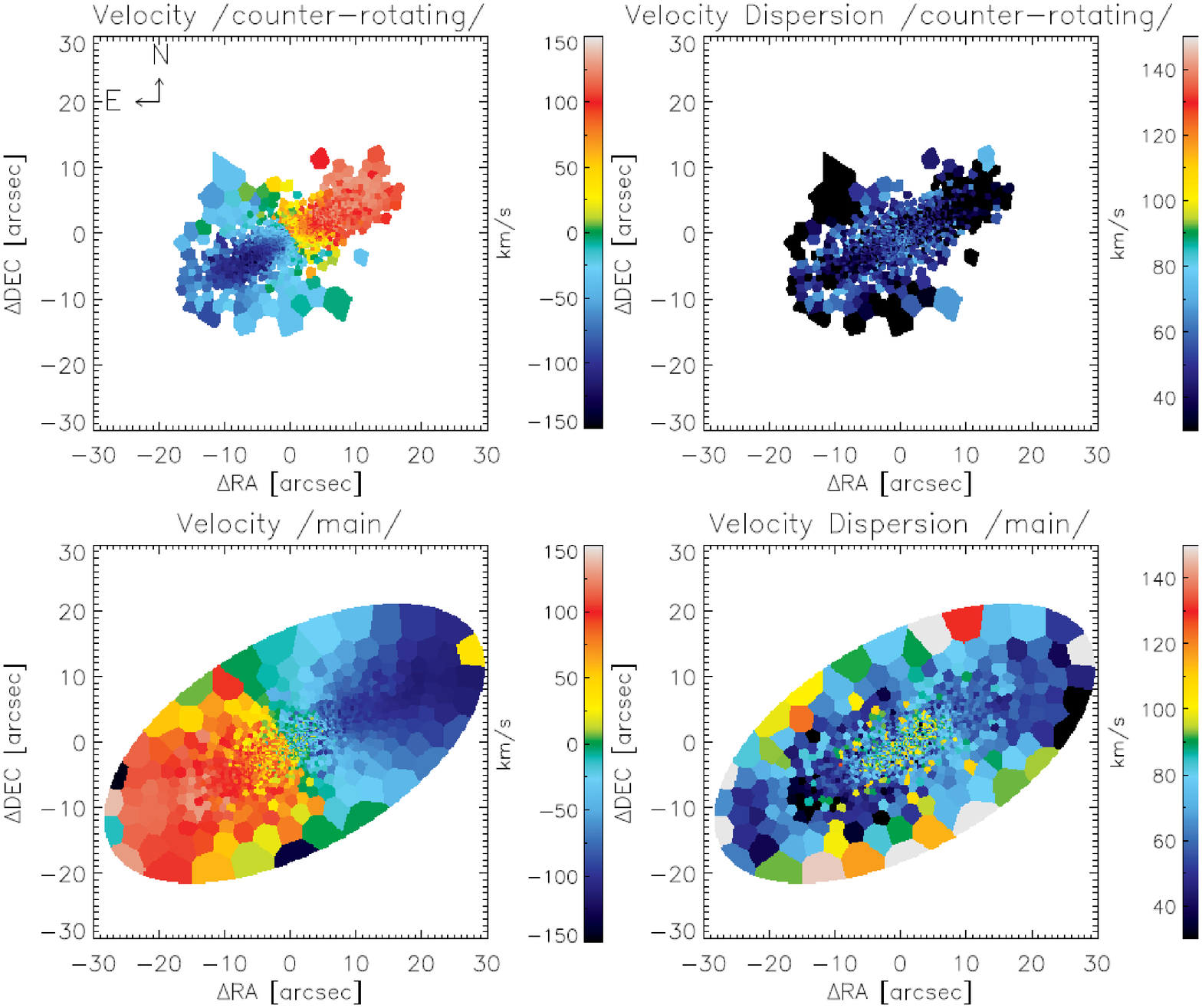}
\caption{The extracted kinematic maps without imposing any priors on the fractional light contribution for NGC~448. The bins are selected such that the fractional contribution of the counter-rotating component is greater than 15$\%$. Top-left panel: The velocity field of the counter-rotating kinematic component. Top-right panel: The velocity dispersion map for the counter-rotating component. Bottom-left panel: The velocity field of the main kinematic component. Top-right panel: The velocity dispersion map for the main component.}
\label{fig:V_MAP_NGC448} 
\end{figure}
}  

\newcommand{\IncludeFigEightt}{
\begin{figure}
\centering 
\includegraphics[width=0.49\textwidth]{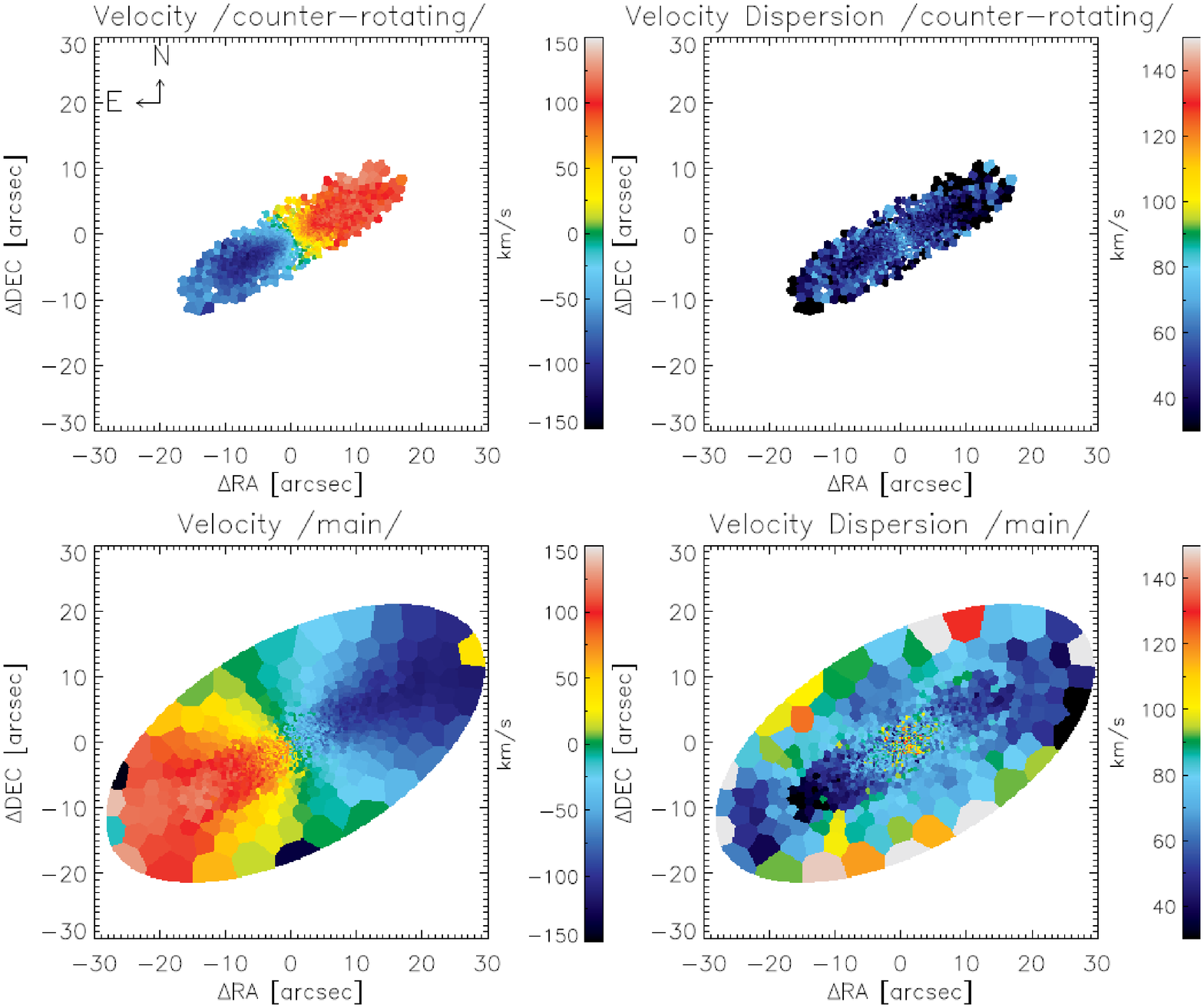}
\caption{The extracted kinematic maps with the fractional light contribution fixed to correspond to the photometrically derived one for NGC~448. The bins are selected such that the fractional contribution of the counter-rotating component is greater than 15$\%$. Top-left panel: The velocity field of the counter-rotating kinematic component. Top-right panel: The velocity dispersion map for the counter-rotating component. Bottom-left panel: The velocity field of the main kinematic component. Top-right panel: The velocity dispersion map for the main component.}
\label{fig:V_MAP_FIX_NGC448} 
\end{figure}
}

\newcommand{\IncludeFigElev}{
\begin{figure*}
\centering
\includegraphics[width=0.99\textwidth]{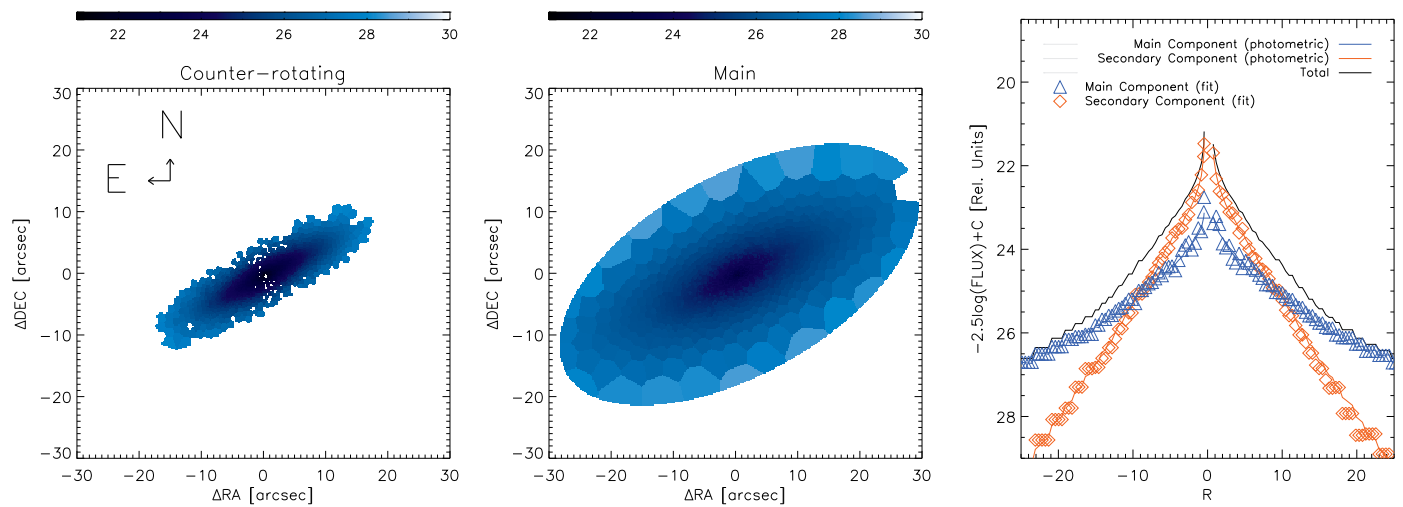}
\caption{Spectral flux maps averaged over the number of spaxels and profile fixed to the one recovered by the photometric decomposition for NGC~448. Left panel: Reconstructed surface brightness map of the counter-rotating component. Middle panel: Reconstructed surface brightness map of the main component. Right panel: Radial profile for both kinematic components, taken through a pseudo slit approximately aligned to the semi-major axis. The black, orange, and blue solid lines illustrate the photometrically derived total, counter-rotating, and main component, respectively, flux contribution. The blue triangles and the orange diamonds show the flux contribution derived of the counter-rotating and main components as derived through our fitting procedure without imposing any constraints.}
\label{fig:IMG_FRAC_NGC448_F} 
\end{figure*} 
} 

\newcommand{\IncludeFigForteen}{
\begin{figure}
\centering
\includegraphics[width=0.46\textwidth]{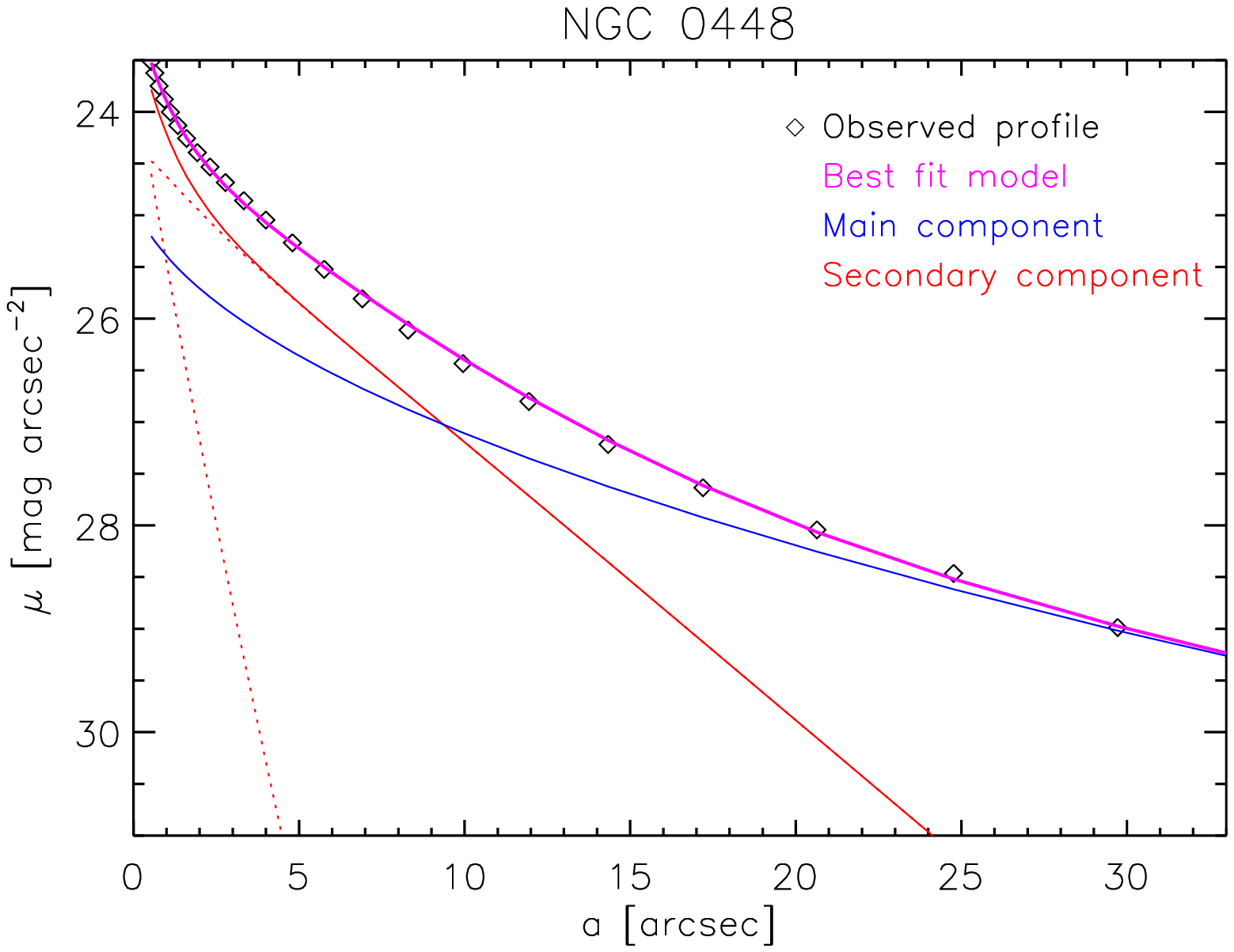}
\newline
\newline
\includegraphics[width=0.46\textwidth]{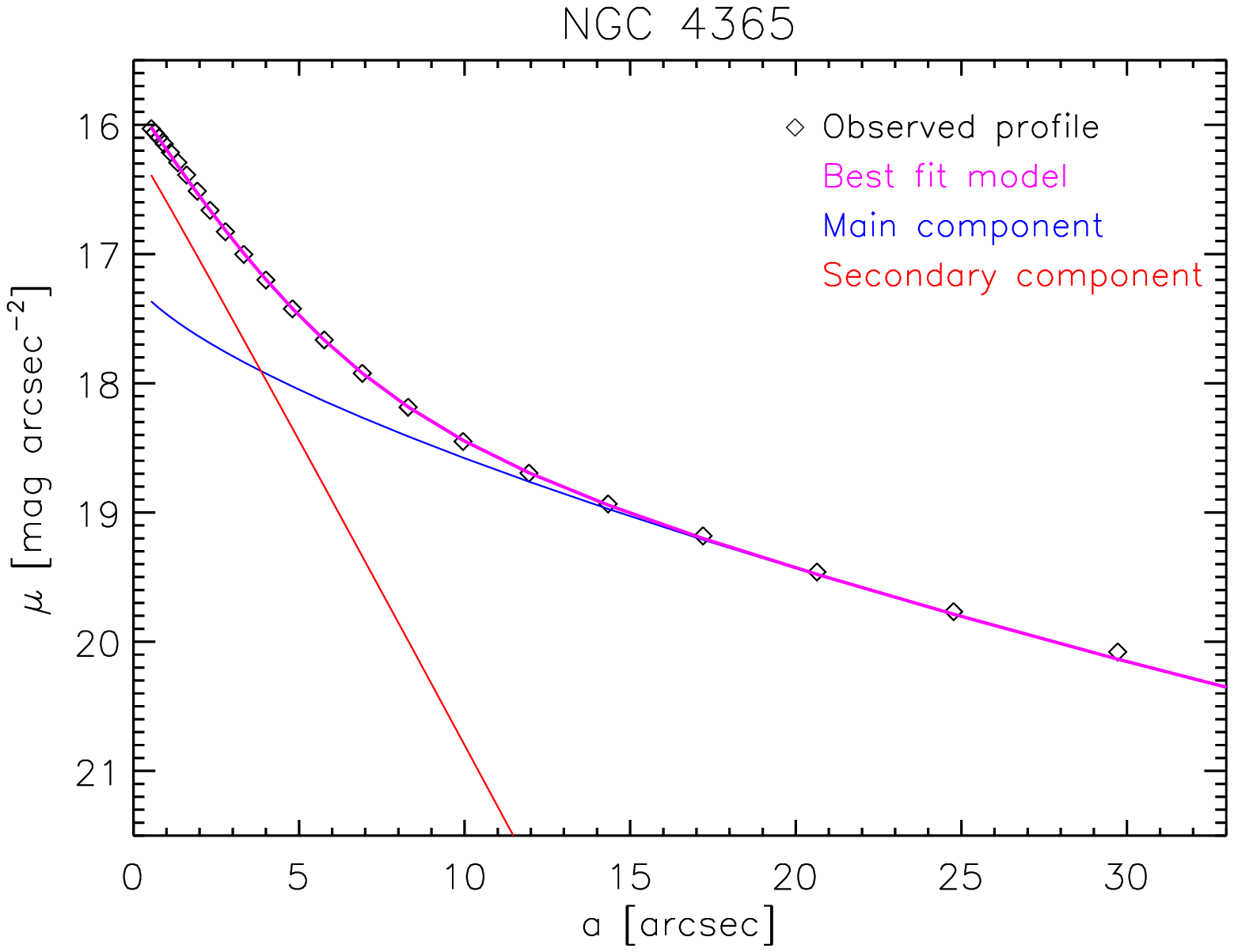}
\caption{Photometric decomposition for NGC~448 (upper panel) and NGC~4365 (lower panel). Each plot shows: the surface brightness profile measured from the reconstructed images (diamonds), the best fit model (magenta) as obtained from the combination of the structural components that are associated to the main galaxy (blue) and to the kinematically-distinct core (KDC, labelled as “secondary component”, in red). In the case of NGC~448 the KDC is assumed to include two subcomponents (dashed red lines).}
\label{fig:photometry} 
\end{figure}
}  

\newcommand{\IncludeFigFifteen}{
\begin{figure*}
\centering
\includegraphics[width=0.99\textwidth]{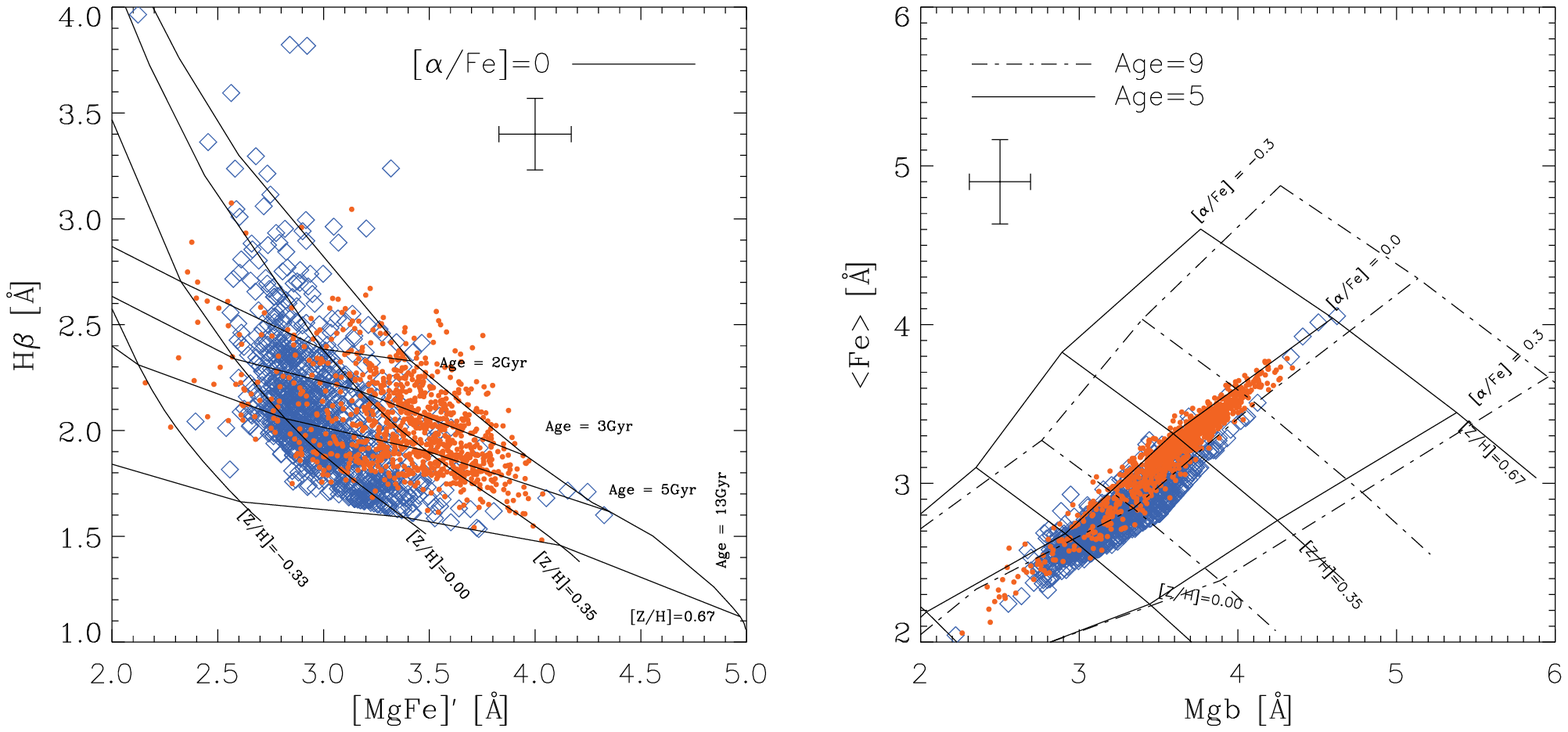}
\caption{Equivalent width of the composite Lick indices for NGC~448 for all bins where the two kinematic components have velocity amplitudes higher than 30 \kms. The \citet{Tho11} model prediction is over-imposed. The blue diamonds signify the main component and the red filled circles indicate the counter-rotating one. The crosses show the mean error.}
\label{fig:ind_diag} 
\end{figure*}
}  

\newcommand{\IncludeFigsixteen}{
\begin{figure*}
\centering
\includegraphics[height=0.85\textwidth, keepaspectratio]{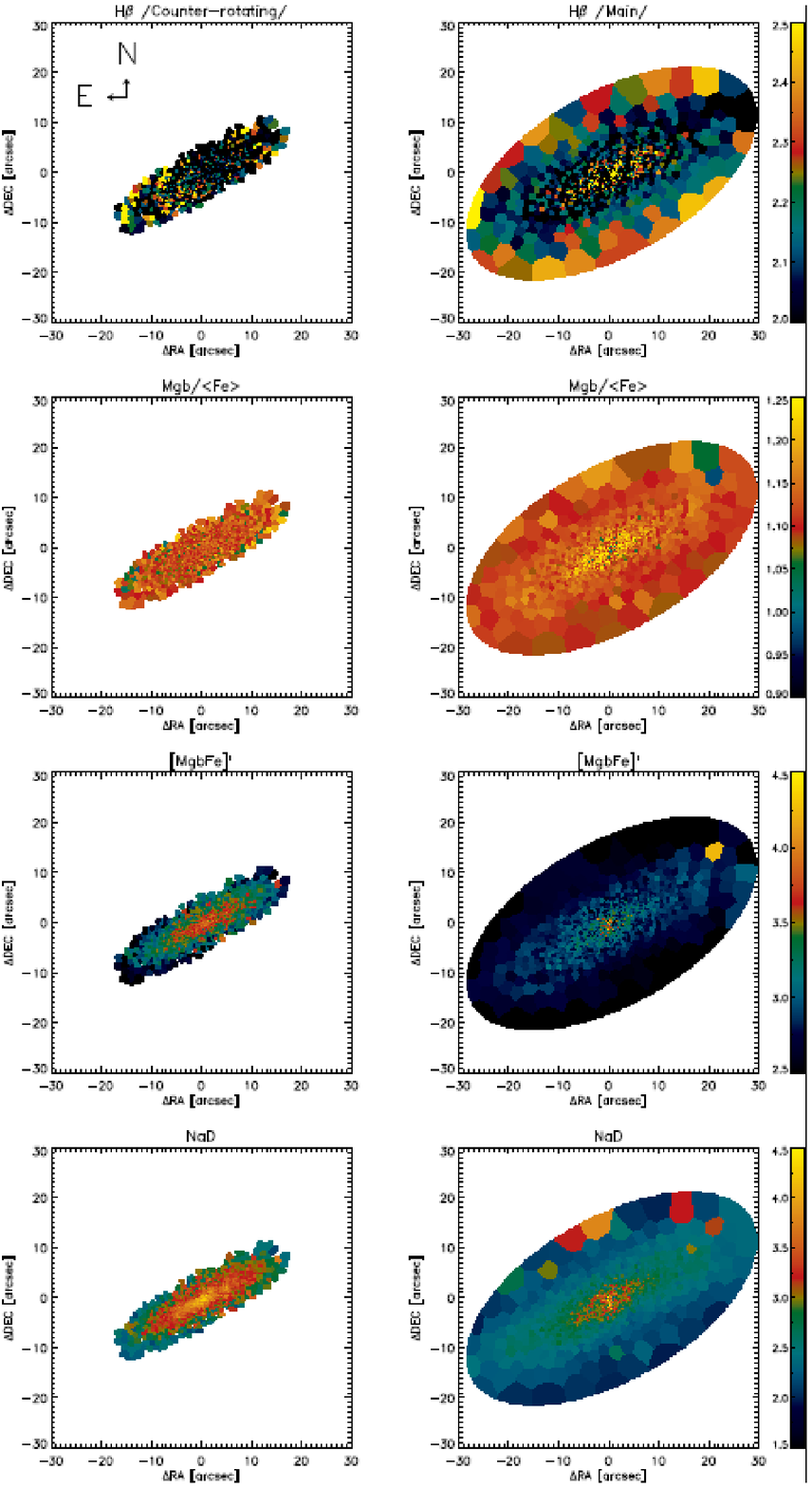}
\includegraphics[height=0.84\textwidth, keepaspectratio]{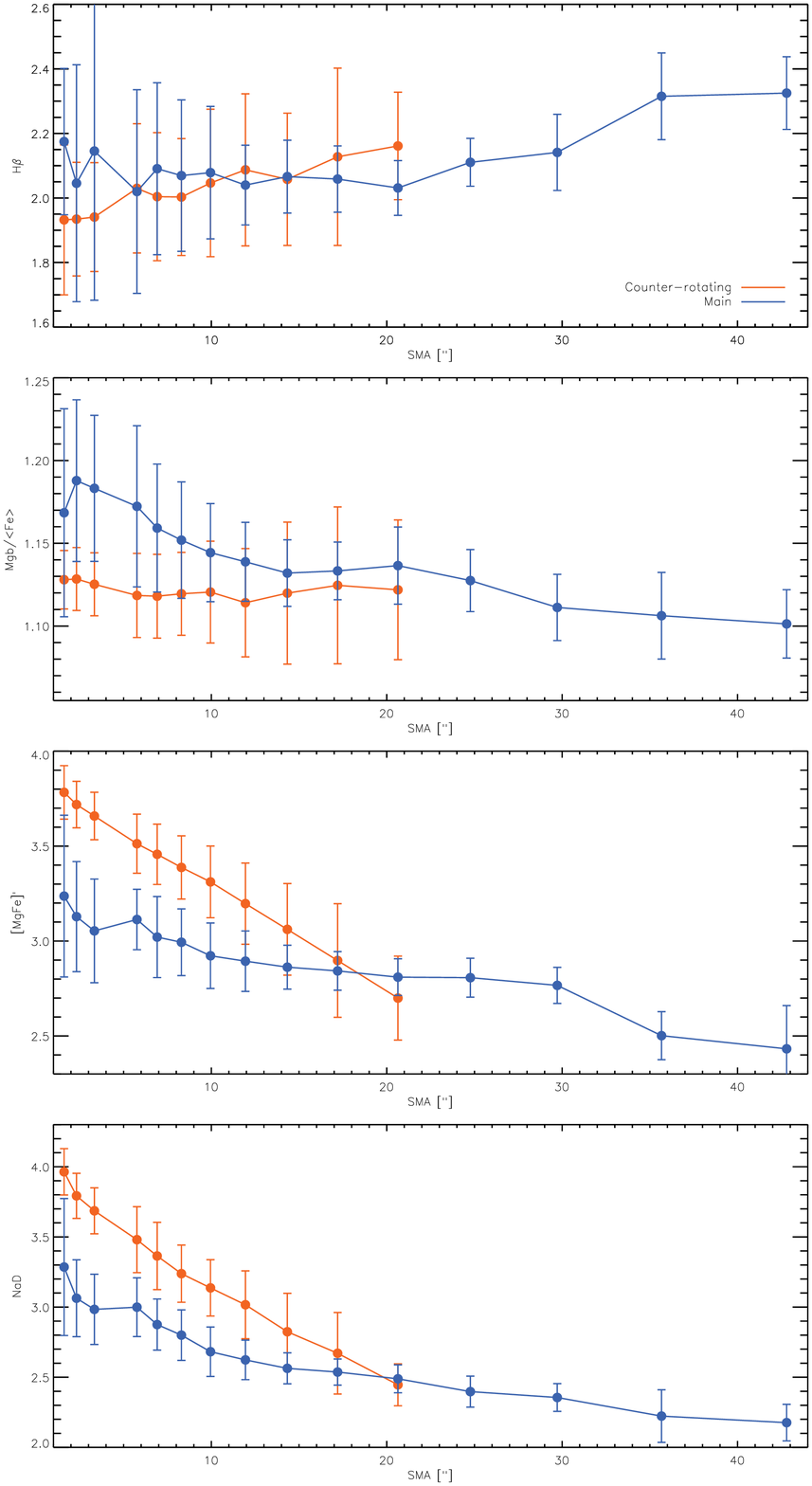}
\newline
\caption{Line-strength index maps and radial profiles for NGC~448. Left panels: Two-dimensional maps of the index strengths of the counter-rotating component of \hb, \mgb/\Fe, \MgFep, and Na D from top to bottom respectively. Middle panels: Same two-dimensional maps of the index strengths, but for the main component. Right panels: The index radial profiles taking elliptical annuli as for the counter-rotating (red) and main stellar (blue) kinematic components. The error bars represent the resistant standard deviation of the measurements within a given annulus.}
\label{fig:ind_m_p} 
\end{figure*}
}  

\newcommand{\IncludeFigsevteen}{
\begin{figure*}
\centering
\includegraphics[height=0.6\textwidth, keepaspectratio]{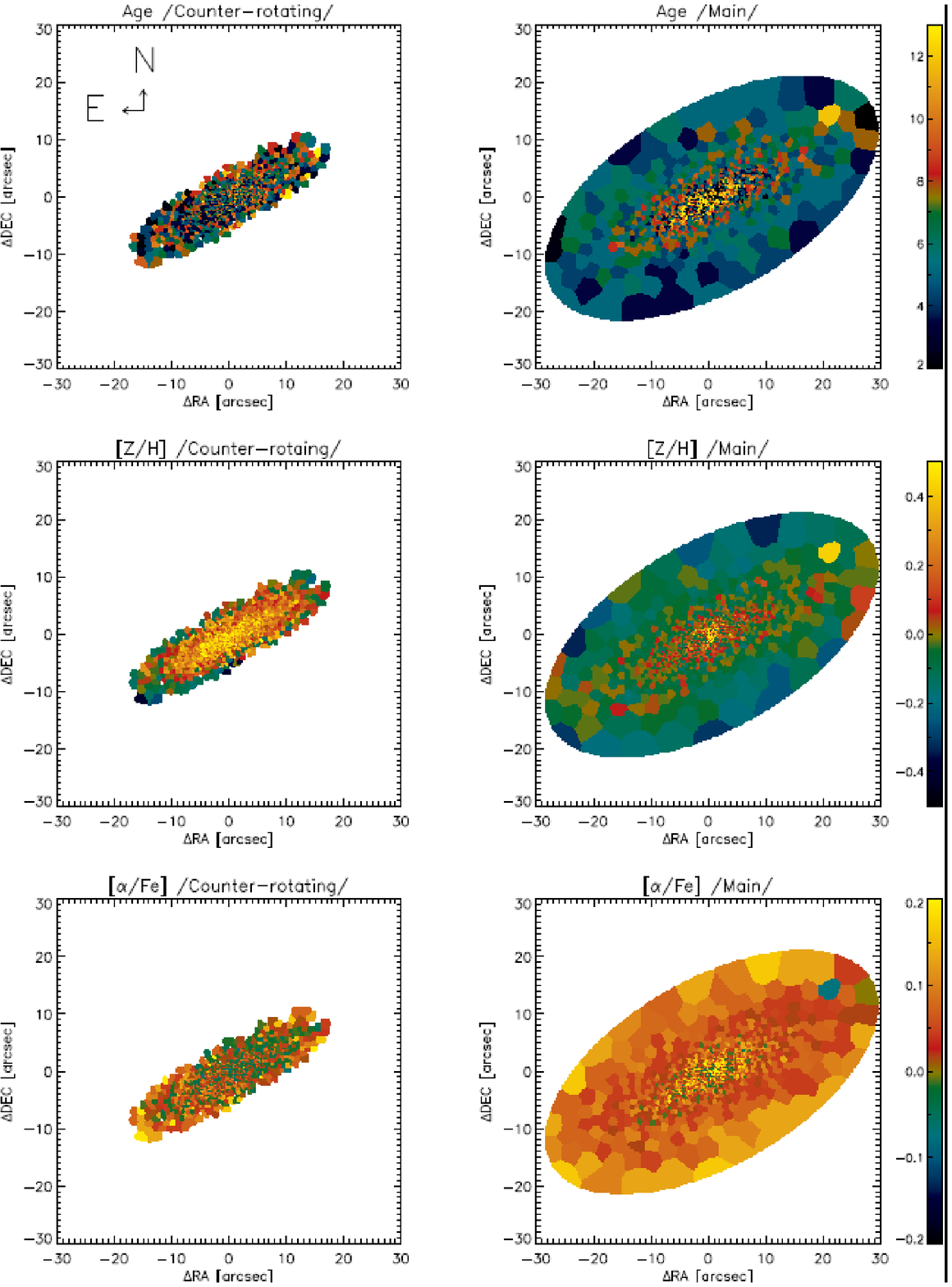}
\includegraphics[height=0.6\textwidth, keepaspectratio]{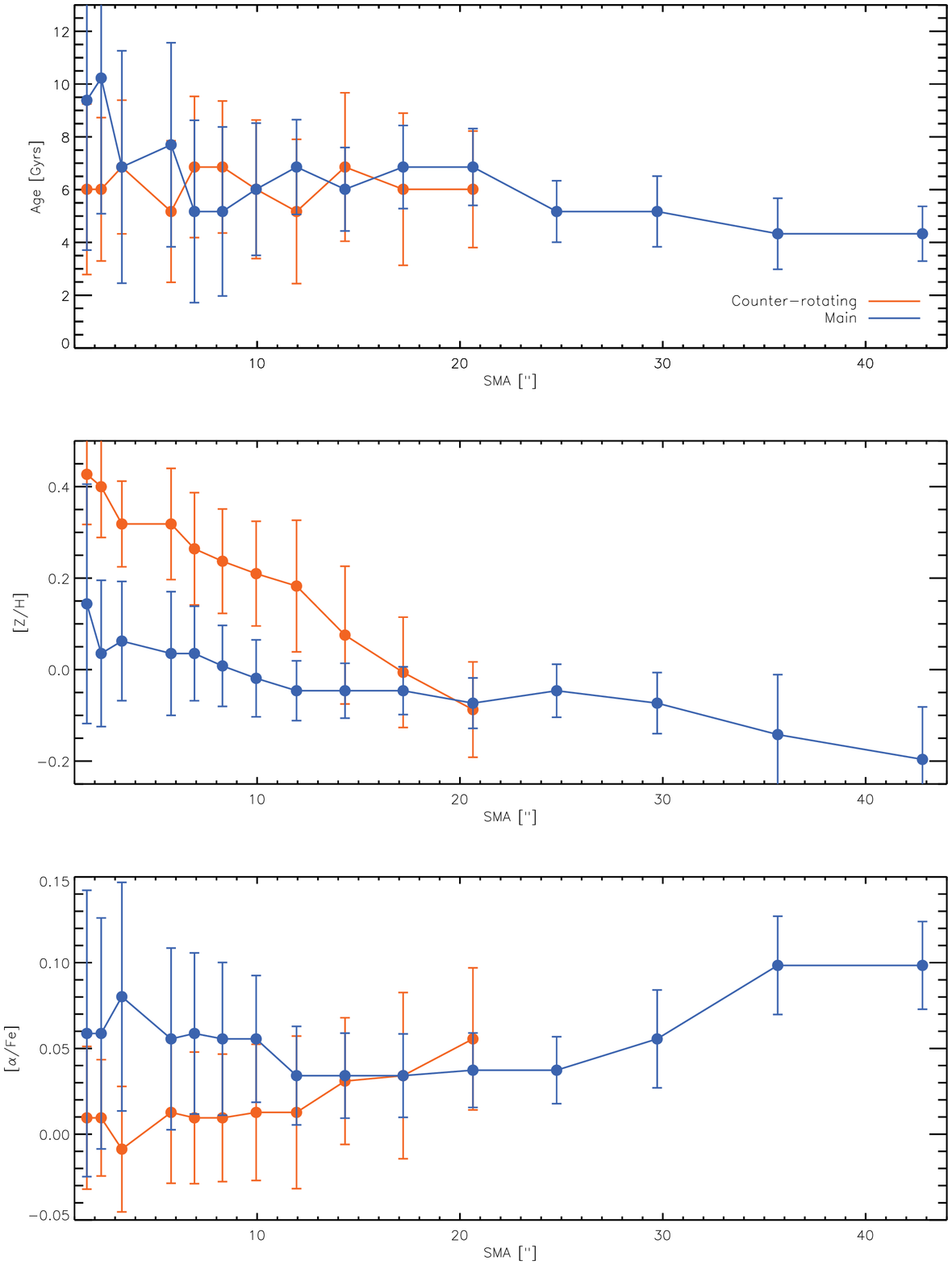}
\newline
\caption{Single stellar population properties for NGC~448 recovered by fitting the Thomas models. Left panel: Two-dimensional maps of the age, metallicity ([Z/H]), and $\alpha$-element-to-iron abundance ([$\alpha$/Fe]) for the counter-rotating component. Middle panel: Same age, metallicity, and [$\alpha$/Fe] maps for the main stellar kinematic component. Right panel: Radial profiles of the best-fitting ages, $[Z/H]$, and [$\alpha$/Fe] of the counter-rotating kinematic component (red) and the main one (blue). The error bars were evaluated by taking the resistant standard deviation of the values within a given annulus.}
\label{fig:a_m_a} 
\end{figure*}
}  

\newcommand{\IncludeFigFo}{
\begin{figure}
\centering 
\includegraphics[width=0.46\textwidth]{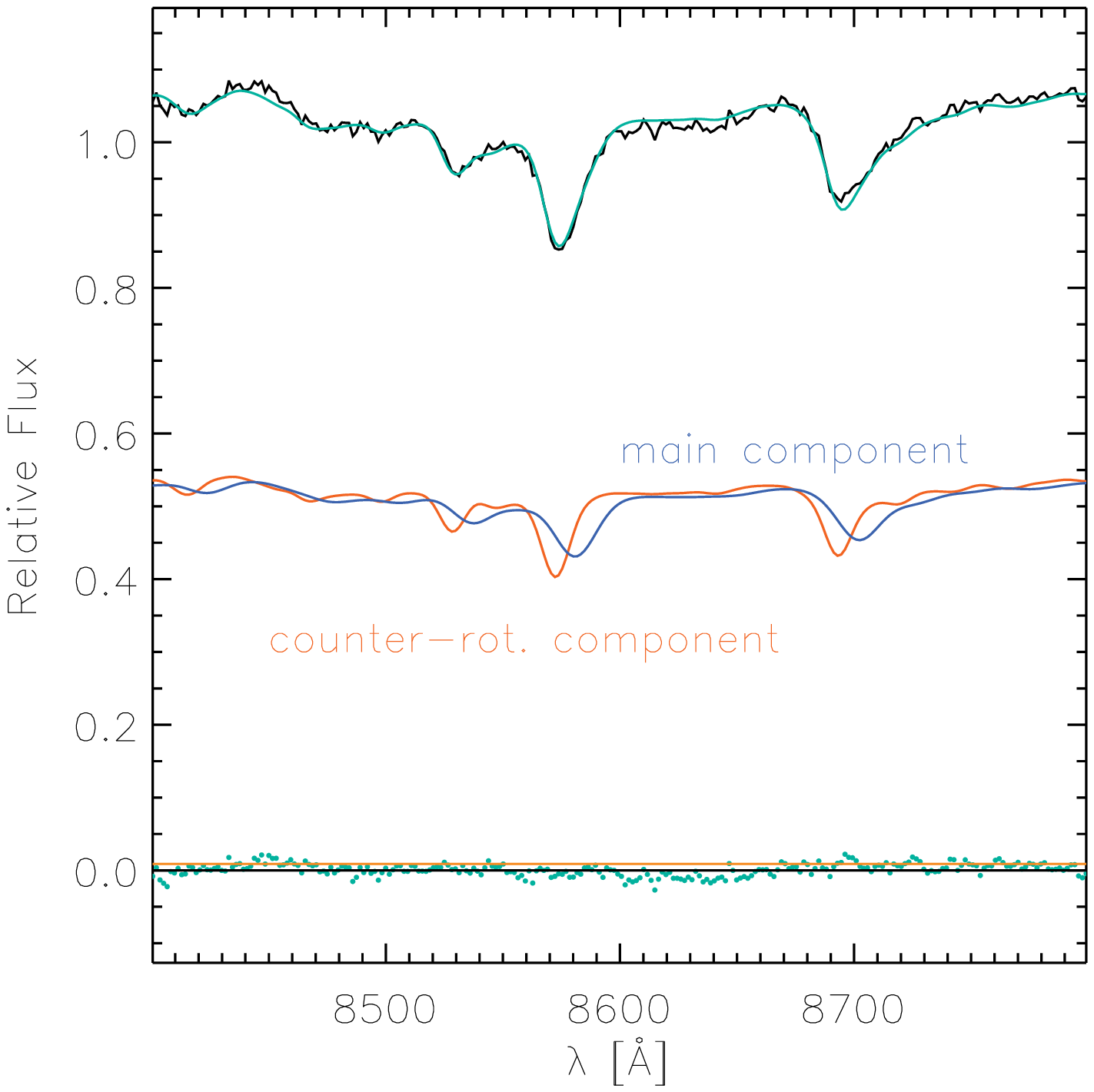}
\caption{Fit to the CaT spectral region of NGC~4365 for a bin approximately 10.5 arcsec away from the field center in the region where we expect a large velocity difference between the two suspected kinematic components.} %
\label{fig:FIT_NGC4365} 
\end{figure} 
} 

 
  \abstract
   {}
   {We study the kinematically distinct components in two early-type galaxies NGC~448 and NGC~4365 aided by integral-field observations with the Multi Unit Spectroscopic Explorer (MUSE) on the Very Large Telescope. NGC~448 has previously been shown to host a counter-rotating stellar disk. NGC~4365 harbours a central (apparently) decoupled core that has been suggested to not be physically distinct from the main body and instead stems from the different orbital types in the core and main body due to its triaxial nature. We aim to measure the brightness profiles, kinematics, and stellar population properties of the peculiar kinematic structures in these galaxies and shed light on their true nature and formation mechanism.}
   {We use a kinematic decomposition technique to separate the individual contributions to the spectra of the two distinct kinematic components observed at each spatial position in the field of view. Furthermore, by folding back the outcome of a photometric decomposition we reduce the intrinsic degeneracies in recovering the kinematics and the best-fitting stellar spectral templates. Finally, by extracting the Lick line-strength indices for the individual components and fitting them to single stellar population models we derive their ages, metallicities, and $\alpha$/Fe overabundances.} 
   {The two kinematically decoupled stellar components in NGC~448 have similar ages, but different chemical compositions. The distinct kinematic feature in NGC~448 has a nearly exponential surface-brightness light profile, dominates in the innermost $\sim 10\arcsec$, is smaller in size, and is very likely an embedded counter-rotating disk as also indicated by its kinematics. It has higher metallicity than the main galaxy stellar body and lower $\alpha$/Fe overabundance. By contrast, we do not find evidence for true decoupling in the two distinct kinematic components in NGC~4365. This confirms earlier work suggesting that the kinematically distinct core is likely not a separate dynamical structure, but most certainly likely a projection effect stemming from the orbital structure of this galaxy that was previously found to be intrinsically triaxial in shape.}
   {Our findings indicate that the kinematically decoupled component in NGC~448 is truly decoupled, has external origin, and was formed through either the acquisition of gas and a subsequent star-formation episode or it formed from the direct accretion of stars from a companion. Conversely, the presence of a kinematically distinct component in NGC~4365 is not associated to a true kinematic decoupling and is instead most likely due to projection effect stemming from the triaxial nature of this galaxy.} 
\keywords{galaxies: abundances - galaxies:individual: NGC~448 - galaxies: individual: NGC~4365 - galaxies: kinematics and dynamics: galaxies: stellar content - galaxies: elliptical and lenticular, cD}

\titlerunning{The kinematically distinct components in NGC~448 and NGC~4365}
\authorrunning{Nedelchev et al.}
\maketitle
%

\section{Introduction}
\label{sec:Intr}

The kinematic and stellar populations properties of local galaxies carry the imprint of their formation and evolution histories. Within the local Universe most galaxies display rather regular kinematic appearance \citep{Kraj08, Cap16}. However, a fraction of the local galaxies show some peculiarities in their stellar \citep[e.g.,][]{Kraj11} and gas \citep[e.g.,][]{Sar06} kinematics or both. Often these peculiarities imply underlying rotation around more than one axis (or anti-parallel spin), comprising a class of galaxies labelled as 'Multi-Spin Galaxies' \citep{Rub94}. The observed multiple spin axes generally could arise from a number of morphologically different structures. These range from polar ring or disk structures \citep[e.g. NGC 4650A,][]{Iod15}, large-scale counter-rotating disks \citep[e.g. NGC 4550,][]{Cocc13} to kinematically distinct or counter-rotating cores \citep[e.g. NGC 5813,][]{Kraj15}. The latter population of galaxies with a kinematically distinct core (KDC) displays an abrupt change in the direction of the velocity field accompanied by a large change in the kinematic position angle. In some of these special cases the velocity of the KDC switches sign (i.e. the two kinematic components show opposite rotation to one another) to differentiate a subclass of KDC that are specifically defined as counter-rotating cores (CRC). Another signature of a counter-rotating component is the presence of two symmetric velocity dispersion peaks along the main rotation axis in a subclass of multi-spin galaxies labelled as ``$2\sigma$'' \citep{Kraj11}. \textit{Intrinsic} counter-rotation is observed when the rotation occurs around the same rotation axis, but involves different kinematic components rotating in opposite directions. Conversely, if the (decoupled) components have different misaligned rotation axes and their angular momentum vectors simply project onto the sky plane anti-parallel, the counter-rotation is deemed to be \textit{apparent} \citep{Cor14}. 

Observationally, a number of disk galaxies (i.e. some lenticulars and spirals) have been demonstrated to possess a truly decoupled kinematic structure that counter-rotates with respect to the main stellar body and often has a disk-like morphology \citep[e.g][]{Cocc11,Cocc13,Cocc15,Kat13,More17}. Either through long-slit or integral-field unit (IFU) spectroscopic observations it has been inferred by utilising a spectroscopic decomposition technique \citep{Cocc11} that the distinct cores of such galaxies are more extended embedded components, comprise younger stellar populations, and are always accompanied by a substantial gas reservoir traced by ionised gas emission \citep{John13,Pizz14,Mitzk17}. 

Owing to their peculiar nature, galaxies with counter-rotating components have been investigated with the aid of numerical simulations. Several different formation scenarios have been proposed and modelled to explain their existence. The most favoured formation channel for such kinematically distinct components in lenticulars and spirals is thought to be the external acquisition of gas as a consequence of a gas rich major or minor merger and a subsequent period of in situ star formation \citep{Pizz04}. Detailed numerical modelling has indeed shown that both gaseous and stellar counter rotation can result from a minor merger scenario, provided that initially the two progenitors have opposite spins and that the gas content of one of the progenitors is substantially higher than that of the other such that the gas accreted on retrograde orbits is not dissipated and swept up by any pre-existing gas \citep[e.g.,][]{Bass17}. Another channel for the formation of disk-like counter-rotating distinct kinematic components is proposed in \citet{Alg14}. The consequential gas accretion from two distinct large-scale cosmological filaments under a very specific spatial configuration in their zoom-in cosmological simulation naturally forms a population of counter-rotating stars implanted within the host galaxy. \citet{Eva94} also suggested that counter-rotating stellar disks can form through the internal separatrix crossing mechanism. The change from elliptical to circular disk (e.g., through the disbandment of a bar structure in a triaxial halo) shifts some stars, initially moving on box orbits across the separatrix to tube orbits, shaping two co-rotating and counter-rotating stellar populations. Another mechanism that was put forward to explain the formation of two counter-rotating disks is the (near) coplanar major merger of two opposite spin spirals \citep[e.g.,][]{Pue01,Cro09}. In such a merger the resultant prograde rotating stellar disk gets ``heated'' more than the retrograde one to naturally explain the kinematic structure of galaxies similar to NGC 4550 \citep{Cro09}.

The situation is even more complicated with regards to ellipticals. KDCs have long been known to occur frequently in the centres of elliptical galaxies \citep[e.g.,][]{Ben88}. Generally, distinct kinematic features can be the tell-tale signs for either intrinsically more centrally concentrated embedded small components, or just ``top-of-the-iceberg'' indication of more extended kinematic structures. Nevertheless, \citet{McDer06} reported a dichotomy in the KDCs found within massive ellipticals. The elliptical slow rotators mainly host more extended KDCs (on kpc scales) with stellar populations of age comparable to that of the stars within the main galaxy body. Conversely, the more centrally concentrated KDCs (of a few hundred pc) usually have noticeably younger stellar populations and are predominantly found in fast-rotating galaxies. The KDCs in massive ellipticals could also be a mere projection effect. The (weak) triaxial nature of some of these galaxies \citep{Cap07} allows them to support multiple types of different orbit families \citep[e.g.,][]{deZ91} and therefore viewed from a different vantage point these can appear to possess a KDC, nicely illustrated in \citet{Sta91}. 

A lot of effort from a simulation standpoint has also been focused towards understanding the nature of the kinematically distinct components within more massive ellipticals. Ever since their first detection these KDCs have been ascribed to be the signature of a past merger \citep[e.g.,][]{Kor84, Balc90}. 
Indeed, \citet{Hoff10} proposed that a merger event involving progenitors harbouring a non-negligible gas reservoir ($\sim$ 20\%), can frequently result in elliptical galaxies that exhibit a KDC whose properties resemble that of a faster (counter-) rotating embedded disk. \citet{Bo11} performed another comprehensive set of high-resolution merger simulations with various initial progenitor mass ratios. Their results showed that the merger remnants of an opposite spin spiral galaxy pair resemble slow rotators, often with a central KDC, albeit not reproducing all the morphological and kinematic properties of massive elliptical galaxies in the ATLAS$^{\rm 3D}$\ survey \citep{Cap11}.

Moreover, KDCs need not be just a consequence of a merger of opposite spin galaxies. The reactive force stemming from the substantial mass loss in the merger process can act to naturally place gas or stars or both on retrograde orbits and conspire to form a KDC embedded in the remnant elliptical galaxy \citep{Tsat15}. Most recently, \citet{Sch17} followed the formation and evolution of a small-scale (and almost certainly therefore younger) KDC in a high resolution equal mass merger simulation with specific emphasis on the KDC stability. Perhaps surprisingly, they discovered that such structures could be semi-stable for about 3 Gyr and highly dynamic, undergoing global gyroscopic precession, before being gradually dissolved and dispersed within a few Gyr.    

In this paper, we study the kinematically distinct components in NGC~448 and NGC~4365 with the aim to disentangle their kinematics, stellar population properties, and morphologies. These early-type galaxies were selected from the ATLAS$^{\rm 3D}$\ survey as part of an effort to address the nature of kinematically distinct components. They were chosen to be the hosts of pronounced kinematically distinct component. The visible apparent size (diameter) of the kinematical peculiarities in previous observations of these two galaxies was selected to be more than $10\arcsec$ to further safeguard against potential complications that can arise in trying to apply our kinematic decomposition technique and ensure its robust performance.
\citet{Kat16} independently studied NGC~448 and extracted a long-slit spectrum along the galaxy major axis to perform a thorough investigation of the properties of the two kinematically decoupled components. Our second galaxy NGC~4365 was also extensively observed with the \texttt{SAURON} integral-field spectrograph \citep{Dav01} and comprehensive dynamical modelling by means of the \citet{Schw79} orbit superposition method has shown that the KDC is not necessarily a true kinematically decoupled structure, but instead appears to be an artefact due to the triaxial nature of NGC~4365 \citep{vdBos08}. 

This paper is organized as follows. In Section~\ref{sec:ObsR} we describe the observations and detail the performed data reduction. Section~\ref{sec:PhDec} outlines our photometric analysis of the two galaxies. In Section~\ref{sec:Kin} we describe the method by which we obtained the one-component kinematics and the resultant kinematics for both galaxies. Then we provide a short summary of the kinematic decomposition technique and present the outcome of such a decomposition. Section~\ref{sec:LSISSP} presents the results of our absorption-line strength measurements and the results of our single-stellar population modelling for the individual kinematic components of NGC~448. Finally, in Section~\ref{sec:Dis} we discuss our finding in light of the possible formation scenarios of the KDCs in these galaxies and summarise our conclusions.
\section{Observations and data reduction}
\label{sec:ObsR}

The spectroscopic observations were carried out with the MUSE integral field spectrograph mounted on the UT4 telescope of the Very Large Telescope at the Paranal Observatory (Chile). Observations were taken on November-December 2014 (NGC~448) and on 12 February 2015 (NGC~4365)  in service mode during grey time and photometric conditions. The seeing was about 1\arcsec, as measured from the ESO meteo monitor. MUSE was configured in wide-field-mode, without adaptive optics, and in nominal wavelength mode; this setup ensured a field of view of $1\arcmin \times 1\arcmin$, a spatial sampling of $0 \farcs 2 \times 0 \farcs 2$ per pixel, a spectral coverage of 4800 -- 9300 \AA\, and a spectral sampling of 1.25 \AA\ pixel$^{-1}$. 

The observations were organised in several exposures of 1260 s intervals with offset sky exposures of 270 s in between. Scientific exposures were dithered by $\sim 1\arcsec$ and rotated by 90$^{\circ}$\ with respect to each other to minimise the instrumental signature of the different spectrograph on the combined data cubes. The total exposure times are 42 m and 2.9 h for NGC~4365 and NGC~448, respectively.

The basic data-reduction (bias subtraction, flat fielding, and wavelength calibration) was carried out using the MUSE ESO pipeline version 1.6.2 \citep{Weil12}. 
We also utilised the Zurich Atmosphere Purge (ZAP) sky subtraction procedure \citep{Sot16b} to remove any residual sky features left over by the main MUSE ESO data reduction pipeline. The ZAP code implements a sky subtraction method based on principal component analysis complemented with data filtering and segmentation that allows robust flux preservation with negligible or no impact on any astronomical source line profiles. These two procedures were executed through their ESO Reflex environment \citep{Freud13} implementations. 
In the case of NGC~448 the ZAP cleaning was done by directly evaluating the sky on the object cube itself. For NGC~4365 we used the dedicated sky exposures as the galaxy occupied the whole field of view, therefore leaving no spaxels dominated purely by the sky signal.
As a subsequent step we ran the \citet{Cap03} Voronoi binning algorithm aiming to achieve a signal-to-noise ratio (S/N) of $90$, evaluated considering the whole MUSE spectral wavelength range. We consider the mean of the spectrum as the signal and the noise as the square root of the mean of the variance returned by the MUSE ESO pipeline. 
Furthermore, to suppress any of the low-signal levels in the outer parts of the observed field of NGC~448 we chose to apply a further elliptical mask aligned to the galaxy position angle (PA) with a semi-major to semi-minor axis ratio of 1.5 in pixel space on the MUSE reconstructed white-light image with a semi-major axis of 190 spaxels, therefore, minimising the effect of the outer regions with very low S/N to the size of the bins in sections of higher S/N. No additional masking was used in the case NGC~4365. 

\section{Photometric decomposition}
\label{sec:PhDec}

\IncludeTblone
\IncludeFigForteen

In this section we exploit the exquisite imaging capabilities of MUSE by analysing the galaxy images obtained by collapsing the datacubes along wavelength. The purpose is to identify the presence of photometrically distinct components and compare their properties with those of the structural components that we would identify via the spectroscopic decomposition (see \S~\ref{subsec:TKin}). This way, it is possible i) to establish a direct association between photometric and kinematic components; and ii) to use the photometric information as priors in the spectroscopic decomposition to constrain the flux of the components. The use of photometric priors in the spectroscopic decomposition to improve the accuracy of the kinematics and stellar population measurements has been successfully applied in several cases \citep[e.g.,][]{Cocc14, Cocc15, Sar16, Tab17}.

The photometric decomposition is performed using the two-dimensional fitting algorithm {\tt galfit} \citep{Peng02}. We use the reconstructed images rather than independent archive images in order to have the same conditions (e.g.,seeing, sky background contamination) as the spectroscopic data. Indeed, the purpose is not to retrieve accurate parameters of the structural components (e.g., seeing-corrected scale radii), but to obtain good fitting surface brightness profiles that contain the same observational effects as the spectra, in order to eventually use them as constraints in the spectral decomposition.

In our decomposition all components are parametrised with Sersic profiles, which allow to encapsulate the behaviour of both disk and spheroidal structures. Our fits include only the minimum number of components required to reach a good fit, rejecting additional components that would contribute only a relatively small fraction of the stellar light (e.g. less than the 5\% freedom that we already allow in our spectroscopic decomposition when adopting photometric priors, \S~5.1).  For NGC~448 we identified 3 Sersic components and for NGC~4365 we identified 2. We also note that fits based on one Sersic component never lead to satisfactory models for the observed surface-brightness distribution, in particular in the regions where we observe a KDC signature. Table 1 summarises the properties of our identified components, whereas in Figure 1 we compare their surface brightness radial profiles with those measured on the reconstructed images using the iraf task ellipse \citep{Jed87}.

As regards NGC~448, we already report here that the sum of the first two photometric components in NGC~448 is likely associated with the kinematic counter-rotating component whereas the third photometric component pertain to the main galactic stellar kinematic component. We also note that within the MUSE field of view our photometric model agrees well with the what predicted by \citet{Kat16} on the basis of their innermost first three components, consisting of a Sersic and two exponential profiles. Indeed, even though their fourth and most extended Sersic component adds up to 17\% of the total light of NGC~448, this halo component does not contribute much light within the MUSE field of view, dropping already to less than a 6\% fractional contribution at a radius of 20\arcsec.

As for NGC~4365 we note that although a single core-Sersic profile can match well the light profile in the KDC region of this galaxy \citep{Fer06}, here we performed a two-component decomposition for the purpose of testing whether these would then match the results of a spectral decomposition. As it will turn out we will find no evidence for two structurally different stellar components in NGC~4365.

\section{Stellar kinematics}
\label{sec:Kin}

\subsection{One-component kinematic fit}
\label{subsec:OKin}

As a first step in extracting the individual properties of our two distinct stellar structural components in these galaxies we used the \citet{Cap04} Penalized Pixel-Fitting (pPXF) and the Gas AND emission Line Fitting (GANDALF) \citet{Sar06} methods to recover the velocity, velocity dispersion, and the {\tt h3} and {\tt h4} coefficients of the Gauss-Hermite parametrisation \citet{vdMar93} of the line-of-sight (stellar) velocity distribution (LOSVD) and any potential ionized-gas emission line properties in each Voronoi bin. The pPXF procedure uses a model, which is parametrised in terms of Gauss-Hermite functions for the LOSVD and a set of linearly combined template spectra to best match the observed galaxy spectrum in pixel space. An important requirement for the extraction of an accurate LOSVD is that the spectral resolution of the templates provided to build the observed galaxy spectrum matches the instrumental one. The spectral resolution was measured by fitting high-resolution solar spectrum template to the twilight spectra, after having combined them following the same sequence of the science observations, similar to \citep{Sar18}. The measured instrumental FWHM is $2.8$ \AA, its variation across the field of view and the wavelength range is $\sim 0.1$ \AA. The adopted method ensures the measurement of the effective instrumental spectral resolution, that includes both the instrument properties and data reduction steps. With these considerations in mind we adopted as template spectra the \citet[MIUSCAT]{Vazd12} stellar population synthesis models (SSPs), spanning the broad 3465--9469 \AA\ wavelength range based on the \citet{Gir00} stellar isochrones with -0.71<[Z/H]<+0.22, 0.06<t<18 Gyr, and with an unimodal initial-mass function with a power-law slope coefficient of 1.3 \citep{Sal55}. This stellar population synthesis model template library with a constant FWHM resolution of 2.8 \AA\ was retrieved through the on-line portal\footnote{http://www.iac.es/proyecto/miles/pages/webtools/tune-ssp-models.php}. Prior to performing both pPXF and GANDALF fits, the spectra of both our template library and observations are logarithmically re-binned to a common velocity step of $55.17$ \kms. 
Due to the higher spectral resolution and relatively high S/N of the CaT absorption lines we extracted the kinematics of NGC~448 within the 8350-8900 \AA\ spectral region. A first degree additive polynomial and a third degree multiplicative polynomial were used to account for template mismatch, any imperfections in the sky subtraction procedure, and potential inaccuracies stemming from the spectral calibration process. 
For NGC~4365, we instead opted to extract the kinematics by creating an optimal template to match the summed up spectrum within a central circular aperture by performing an initial pPXF fit. This template and a high-order additive and multiplicative polynomial (15th and 15th degree, respectively) were later adopted to similarly extract the one kinematic component maps through pPXF and GANDALF within each Voronoi bin. 

\IncludeFigOne
\IncludeFigTwo
\IncludeFigThree

Figure~\ref{fig:SPEC_NGC448} presents the two dimensional kinematic maps for NGC~448 extracted by performing the procedure described above.  We observe abrupt changes in the direction of the velocity field (upper-left panel) with a maximum rotation amplitude of $\sim 140$\ \kms.  
There is a pronounced anti-correlation between the velocity and {\tt h3} coefficient in the outer regions of the galaxy plausibly because of the presence of an extended disk component that rotates noticeably different than the main stellar body. Moreover, within the inner central part, this anti-correlation persists even after the velocity field switches direction. We confirm the classification of NGC~448 as a ``2$\sigma$'' galaxy (top right panel), because of the two pronounced symmetric off-centre peaks of $\sim 230$\ \kms\ in the velocity dispersion map. The behaviour of the {\tt h4} Gauss-Hermite coefficient used to capture symmetric departures from purely Gaussian LOSVD profile indicates that at the positions where the velocity dispersion is at its highest values the LOSVD has extended pronounced ``wings''. Taken together, such kinematic behaviour in NGC~448 is found to be symptomatic of the presence of an embedded central intrinsically counter-rotating disk \citep{Rub92,Rix92,Ber96,Cap07,Verg07} that from these maps appears to span at least $\sim 13\arcsec$ in its radial extent.

In Figure~\ref{fig:SPEC_NGC4365} we show the two dimensional kinematic maps for NGC4365. The velocity field (upper left panel) displays a clear ``four-part counter-rotating'' morphology because it possesses four maxima at different amplitudes and positions and the velocity field has a distinctive ``S-shape'' twist \citep{Sta91}. The central velocity extrema have an amplitude of $\sim 80$\ \kms, whereas in the outer regions the velocity amplitude at the extrema is smaller ($\sim 55$\ \kms). The velocity dispersion (upper right panel) is centrally peaked with a maximum value of $\sim 270$\ \kms. The {\tt h3} coefficient (bottom left panel) strongly correlates with the observed velocity pattern where we also observe a kinematically distinct behaviour for the velocity field. Such correlation is less pronounced in the outer parts. 

\subsection{The kinematic decomposition procedure}
\label{subsec:KinDec}

To derive the individual properties of the two different structural components, we followed a procedure identical to that of \citet{Cocc11}. 
This so-called kinematic decomposition method takes advantage of the characteristic spectral imprint (e.g., asymmetric or double trough line profiles) of a spectrum, which constitutes of two overlapping stellar populations of different kinematics. The method builds on pPXF in that it aims to model the observed spectra and therefore to also achieve a separation of two possible stellar kinematic structural components in every spatial bin, by constructing a unique set of linearly combined templates for each component convolving it with a separate Gaussian line-of-sight velocity distribution function.
To account for any potential inaccuracies in the spectral calibration and the effects induced by the reddening from dust on the spectra similarly to pPXF our procedure adopts multiplicative Legendre polynomials of a pre-selected order assumed to be identical for both components. We avoid the use of additive Legendre polynomials, because they could profoundly bias our line strength measurements \citep{Cap17}. 
Given the complexity of this fitting approach, a useful simplification is achieved through the normalisation of both the galaxy spectra and templates to unity. In this way we are able to obtain the fractional contribution to the flux within a given wavelength range of both components in terms of a single parameter instead of assigning two separate light fraction contribution parameters. 
Due to the surprisingly good calibration in both our chosen set of templates and MUSE spectra, we found that such an arbitrary choice of normalising wavelength range does not influence any of the forthcoming conclusions and chose for simplicity to normalise all templates and MUSE spectra with respect to their mean levels. 
The main workhorse minimisation procedure within both pPXF and our kinematic decomposition method is the interactive data language (IDL) implementation \citep{Mar09} of the Levenberg-Marquardt least-squares curve fitting algorithm that is sometimes found to converge, not necessarily towards a global minimum. We experimented by switching it with slower, but potentially more sturdy downhill simplex (AMOEBA) method \citep{Nel65} with little to no success in line with the \citet{Cap17} finding that the minimisation algorithm does not noticeably influence the inferred pPXF results. 
Because even though our fitting procedure is robust enough, it is not necessarily immune to completely unreasonable sets of initial guesses for the velocities of the two kinematic components. We tested the impact of varying the initial guesses for both kinematic components (i.e. their velocities and velocity dispersions) in our decomposition model until we were convinced of the general validity of the results obtained.
\IncludeFigSix
\subsection{Two-component fit}
\label{subsec:TKin}

As described above such a kinematic decomposition procedure allows us to simultaneously and independently retrieve the kinematic components properties, namely their velocity, velocity dispersion, fractional contribution to the flux (light fraction), and best-fitting stellar population template. Here, we focus on the individual photometric and kinematic properties of the observed distinct components of our galaxies. 

\subsubsection{NGC~448}
Figure~\ref{fig:FIT_NGC448} shows an example spectrum in the CaT region for NGC~448 near the position of maximum velocity difference, where  both components contribute almost the same fraction to the overall light (at $\sim 9\arcsec$). 
In the case of NGC~448, the line profiles are clearly double peaked, highlighting the complex nature of the underlying stellar LOSVD. Figure~\ref{fig:IMG_FRAC_NGC448} shows the reconstructed brightness maps averaged over the number of spaxels of both kinematic components.  
The light profiles taken along the photometric major axis shows that our kinematic and photometric decompositions agree remarkably well with each other (right panel of Fig.~\ref{fig:IMG_FRAC_NGC448}), so that the more (primary) and less (secondary) extended photometric components can be confidently attributed to the main and counter-rotating kinematic components, respectively. In this respect, we presently assume that the innermost photometric component belongs to the counter-rotating structure, consistent also with the conclusions of \citep{Kat16}, although our data do not allow to rule out that more central light could be assigned also to the main component. This limitation, however, will not affect our main conclusions. Further adaptive-optics assisted MUSE observations may help in understanding the kinematics of NGC~448 in the very central few arcseconds.

We chose to discard all bins where either the fitting routine did not converge or the fractional light contribution of  the counter-rotating component was less than 15\%. 
This somewhat arbitrary choice is motivated by the observation that the light of the less-extended kinematic component experiences a sharp rise at a radius of $\sim 19\arcsec$ along the pseudo-slit extracted from the IFU data, where its contribution to the total flux indeed drops below 0.15. Furthermore, we retrieved the amplitude of the strongest CaT line and compared it to the level of the residual noise evaluated by computing the robust standard deviation of the residual \citep[eq. 9]{Beer90}. The turn-off where the amplitude of the line drops below three times the residual noise level coincides roughly with our adopted light fraction threshold.  

Nevertheless, because we did not impose any prior on the spatial extent of both of our kinematic components, the decomposition within the regions of small or zero velocity difference in between them,  erroneously detected a higher light fraction in one of the components. Figure~\ref{fig:V_MAP_NGC448} shows the two-dimensional kinematic maps for NGC~448. The upper left panel gives the retrieved velocity for the counter-rotating kinematic component. Both kinematic components rotate with comparable velocities and a maximum velocity amplitude for the counter-rotating component of $\sim 120$\ \kms\ at a distance of $\sim 11\arcsec$ from the centre of the galaxy. Its velocity dispersion is significant lower (mean $\sigma_{\star}$ of $\sim 40$\ \kms) in comparison to that of the main stellar body ($\sim 60$\ \kms), strengthening the evidence that the counter-rotating component is indeed a disk. 
\IncludeFigSeven
\IncludeFigElev
\IncludeFigEightt
Driven by the excellent correspondence between the light profiles of the photometric components determined in \S~\ref{sec:PhDec} and those of the kinematic components, we decided also to impose the light contribution we obtained from photometry into our kinematic spectral model allowing only a small adjustment of at most $5\%$. In Fig.~\ref{fig:IMG_FRAC_NGC448_F} we show again the reconstructed, averaged over the number of spaxels, surface brightness maps of both components and major-axis profiles. We consider again as of our previous estimate that we can reliably detect the counter-rotating component when its contribution to the total flux is higher than 0.15. In the bins where this condition was met (left panel of Fig.~\ref{fig:IMG_FRAC_NGC448_F}) the mean absolute difference between the fixed photometric fractional contribution and the one retrieved by our previous analysis is $\sim 7.5\%$. This difference is higher predominantly within the very central region, where the velocity difference between both kinematic components is rather small and thereby most likely did not originally allow a very robust kinematic decomposition. 
Because of the complex nature of our kinematic decomposition method is not strictly necessary, fixing the photometric contribution relieves some of the parameter degeneracies and results in a more consistent parameter retrieval. Moreover, it eliminates the necessity to impose more fiducial thresholds in separating the Voronoi bins that contain both kinematic components from the ones where just a single one is present than simply adopting a threshold in light fraction.     

We now describe the kinematics obtained by imposing the photometric decomposition, which we adopt in any of our further analysis. Figure~\ref{fig:V_MAP_FIX_NGC448} shows the velocity and velocity dispersion maps for our two kinematically distinct components. We have replaced all bins where we recovered the fractional contribution of the counter-rotating component to be less than 15\% within the main stellar one with our previous single kinematic-component fit. The results that we obtained folding the photometric decomposition information back into our kinematic decomposition also agree up to a high degree with the ones where it was left as a free parameter. The mean absolute differences between the kinematics extracted with and without incorporating the photometric decomposition are $\sim 11$\ \kms\ in velocity and $\sim 12$\ \kms\ in velocity dispersion. The velocity amplitudes for both components are again comparable with the counter-rotating component reaching a maximum of $\sim 125$\ \kms\ at a distance of $\sim 12\arcsec$. The main component has a maximum velocity amplitude of $\sim 135\ $\kms\ in the outskirts of the field along roughly the major photometric axis. The counter-rotating disk component has a lower velocity dispersion and a mean value of $\sim 46$\ \kms\ with a maximum of $\sim 75$\ \kms\ in the very centre and a minimum of $\sim 30$\ \kms. The main component has higher velocity dispersion with a mean value of $\sim 86$\ \kms. 

We report an absence of ionized-gas emission in NGC~448. Although the \oiii, \hb, \nii, and \ha\ emission lines reside in the chosen 4700--6715 \AA\ wavelength range, none of them had a sufficient amplitude-to-noise ($A/N$) retrieved through our GANDALF fit to surpass the detection threshold of $A/N$ > 3. Neither did the \citet{Kat16} long-slit spectrum, nor the \citet{McDer15} ATLAS$^{\rm 3D}$\ observations with \texttt{SAURON} showed the presence of such emission lines. The lack of gas is further enforced by the limits put on the H$_{2}$ mass (logM(H$_{2})$<7.74 [$M_{\sun}$]) of \citet{Young11} based on IRAM CO J = 1--0 and J = 2--1 emission observations. In addition the galaxy is not detected in the FIRST 1.4 GHz survey, excluding the presence of strong enough star formation, that could serve as an ionizing source \citep{Nyl17}. 

\subsubsection{NGC~4365}
\label{subsec:KD4365}
\IncludeFigFo

We also applied the kinematic decomposition technique to NGC~4365. Figure~\ref{fig:FIT_NGC4365} shows a spectrum as of NGC~4365 as example of our attempt at performing such a decomposition in the CaT region. Unlike the case for NGC~448, our kinematic and photometric decompositions did not return light fraction consistent with one another. Moreover, the retrieved kinematics were found to be very highly dependent on the choice of the set of initial guesses. Furthermore, the derived fractional light contribution did not display any consistent profile when left free to vary. Our photometric decomposition shows an excess of light in comparison to a pure S\`ersic fit in the central regions of the galaxy (\S~\ref{sec:PhDec}). The presence of this exponential-like photometric component, however, is most likely due to the structures formed from the orbital families that can be found in galaxies of triaxial intrinsic shape, as previously found for some of the non-barred slow-rotator early-type galaxies with KDCs \citep{Kraj13}, and not a truly kinematically decoupled structure. Even though we obtain a somewhat satisfactory fit to the spectra for some of the Voronoi bins using our decomposition procedure, the velocity and velocity dispersion fields we recover for the two different components are highly irregular. The highly non-symmetric LOSVD profiles in the centres of some ellipticals, such as the one present in NGC~4365, have previously been shown to be well matched by a superposition of two Gaussian forms of different width \citep[e.g.,][]{Fra88,Ben90,Rix92a}. Nevertheless, this finding alone, and especially without a matching photometric signature, does not suffice to provide enough evidence for the presence of a physically decoupled kinematic component.

\section{Line-strength indices and stellar population properties of NGC~448}
\label{sec:LSISSP}
In this section we present the line-strength maps and stellar population properties of the individual kinematic components (i.e. counter-rotating disk and main stellar body) of NGC~448 where our kinematic decomposition technique successfully recovered their presence. 

\subsection{Line-strength indices}
\label{sec:LSI}

Our initial choice to extract the kinematics only in the limited CaT spectral region for NGC~448 did not prevent us from obtaining the Lick absorption line-strength indices as, defined by \citet{Wor94}. We re-fitted the NGC~448 spectra in the 4700--6715 \AA\ wavelength range, where the strongest stellar absorption lines defining the $H\beta$, \mgb, \fei, \feii, \feiii, \feiv, \fev, Na D, \TiOone, and \TiOtwo, using the prior two-component kinematics as an initial guess for our kinematic decomposition. A small adjustment of $\pm 20$ \kms\ in velocity, $\pm 20$ \kms\ in velocity dispersion, and $\pm 0.05$ for the light fraction in the counter-rotating component was allowed to accommodate any expected change in the extracted kinematics. Generally, the kinematics obtained in two sufficiently distant wavelength domains (e.g., 4800--5380 \AA\ and 8480--8750 \AA) could differ. In a varying mixture of different age stellar populations the fraction of young stars, outputting their light predominantly in the blue end of the spectrum, and older stars, dominant at the red end could give rise to non-negligible $\sim 5 $\ \kms\ differences in velocity and velocity dispersion and variation as big as $\sim 0.02$ in the {\tt h3} and {\tt h4} Gauss-Hermite coefficients \citep{Arn14}. This is especially relevant when dealing with possibly kinematically decoupled structures as in our investigation as also observed by \citet{Mitzk17}.
The kinematic decomposition we utilised in modelling the spectrum of our galaxies returns the best-fitting linear combination of library templates. We proceeded to extract the afore-mentioned line-strength indices on them. \citet{Cocc11} carried out extensive verification by means of Monte-Carlo simulations to test the ability of the best-fitting linear combination of templates to capture the underlying stellar population properties under different S/N level and kinematic behaviour albeit with spectra obtained using VLT/VIMOS. The systematic error levels were found to be negligible with respect to the ones driven by the signal noise. We have no reason to believe that the kinematic decomposition procedure would behave differently in our specific case of applying it to MUSE spectra.   
To crudely estimate the error on the measurements of the line-strength indices, we adopted a constant noise level equivalent to $\sqrt{2}$ of the resistant standard deviation ('robust sigma') of the difference between the spectra and our best fit model and the procedure outlined by \citet{Car98}. 
Furthermore, we also evaluated the combined [MgFe]$'=\sqrt{{\mgb} \left( 0.82 \cdot {\fei} + 0.28 \cdot {\feii}\right)}$\ and $\langle{Fe} \rangle$ = $\left( {\fei} + {\feii} \right) /2$  indices \citep{Thom03,Gorg90}. 
Both are considered to be good proxies for the total metallicity. In particular, the [MgFe]$'$ has been demonstrated to be only weakly dependent on the alpha-element-to-iron abundance ratio ($\alpha$/Fe) and the ratio of the \mgb\ to $\langle$Fe$\rangle$ for old stellar populations is found to correlate reliably with the total $\alpha$/Fe enhancement \citep{Thom03}. 

Figure~\ref{fig:ind_diag} shows the strength of these indices for the bins where the relative light fraction of the counter-rotating component is higher than $15\%$ and both components have velocity shifts relative to the galactic systemic one greater than 30\ \kms\ with the corresponding \citet{Tho11} SSP model over-imposed. Within the mean errors the stellar population properties of both kinematic components show discernible differences. The strength of \hb\ for both components is comparable, however, the two kinematic components show a noteworthy difference in their [MgFe]$'$ strengths. The two components are also offset with respect to each other in $\langle$Fe$\rangle$ vs. \mgb\ space with the counter-rotating component having higher values that cover as large a range as seen in the main stellar one, suggestive of a distinction in the $\alpha$/Fe abundance ratios of the two kinematic components. 

These observations are further strengthened by inspecting the two-dimensional maps presented in Fig.~\ref{fig:ind_m_p}. The maps were constructed by taking the values for the main component within the spatial extent where the contribution of the counter-rotating component was identified with the afore-mentioned criteria (i.e. light fraction > 0.15) as our two-kinematic composition and just a single kinematic component fit otherwise. The fact that we observe a smooth transition even after such a combining procedure further validates the basis for our separation. 

The right panels show the profiles of the median values obtained by considering the bins contained within an elliptical aperture defined by performing a fit with the {\tt iraf} task {\tt ellipse} to the reconstructed white image. The velocity amplitude of the kinematic components is taken to be $>30$\ \kms\ as a way to safeguard against any bins where the kinematic decomposition might not have fully converged. The error bars are evaluated by computing the resistant standard deviation within the apertures. At a first glance, both Fig.~\ref{fig:ind_diag} and \ref{fig:ind_m_p} prove that the stellar-population properties of the two kinematic components are indeed different. Moreover, they both show evident radial gradients in [MgFe]$'$, \mgb $/ \langle$Fe$\rangle$, and Na D. The counter-rotating component has lower \mgb $/ \langle$Fe$\rangle$ ratio indicating that its stellar population is less $\alpha$-element enriched. Its metallicity as traced by [MgFe]$'$ is also systematically higher than that of the counter-rotating component. The bottom row in Fig.~\ref{fig:ind_m_p} presents the map and median radial profile for the $\alpha$/Fe insensitive Na D. We notice that this index behaves similarly to the [MgFe]$'$ one. However, due to its higher strength within the main component we could quite clearly separate the presence of a second disk and a halo component, as already hinted by the derived [MgFe]$'$ and expected ubiquitously in S0 galaxies \citep{Gue16}.  

\subsection{Stellar population properties}
\label{sec:StPop}
\IncludeFigFifteen
\IncludeFigsixteen
\IncludeFigsevteen

To translate our line-strength measurements into estimates for the stellar population age, metallicity and alpha-elements abundance for both the main and counter-rotating components, we applied the technique of \citet{Mor12} and \citet{Cocc11} using the single-age models of \citet{Tho11} and the observed strength for the \hb, \mgb, \fei, \feii, \feiii, \feiv, and \fev\ absorption features. Fig.~\ref{fig:a_m_a} presents maps for the age, metallicity and alpha-elements abundance of both stellar components in NGC~448 as well as radial profiles for the median value of these parameters evaluated in elliptical annuli.
The main component is significantly older (median age of $\sim 7.7$\ Gyr) in the very central region (up to $\sim 3\arcsec$ along the major axis) and more metal enriched ([Z/H] of $\sim 0.06$) in comparison to other regions. With this distinction, the ages of the two kinematically distinct components are comparable to each other further outwards. The counter-rotating component is very metal rich (median [Z/H] $\sim 0.15$) and has a steeper gradient with values as large as $\sim 0.4$ falling to as low as $\sim 0$ in its outer parts. On the other hand, the median metallicity of the main component over the extent of the counter-rotating one is $\sim -0.05$. The alpha-enhancement of both components is observed to anti-correlate with metallicity. For the main component it ranges from $\sim 0.05$ in the central region to $\sim 0.1$ outwards. The counter-rotating component shows a hint of a reverse $\alpha$-element abundance gradient. The values are very close to Solar in the centre and rise to match the ones observed for the main component further away.  

Overall our stellar-population measurements for the two kinematic components of NGC~448 do not compare very well with the results of \citet{Kat16}. Along the major axis the stellar metallicity and age derived by \citet{Kat16} are systematically lower and higher, respectively, than our own values, and both in the case of the main and counter-rotating component. It is difficult to comment on these differences, since the analysis of \citet{Kat16} is based on different stellar-population models and analysis technique than ours, namely the NBURST spectral-fitting technique of \citet{Chil07a, Chil07} with the PEGASE.HR SSP models of \citet{LeBorg04}. Furthermore, \citet{Kat16} derive only stellar ages and metallicities whereas we further explore the role of alpha-element abundances.

On the other hand, \citet{McDer15} use a similar modelling approach to ours, based on line-strength measurements, but only consider the overall galaxy spectrum inside circular apertures without attempting to characterise the two stellar populations of NGC~448. Nonetheless, the \citet{McDer15} measurements and results can serve as a test for our modelling approach. Indeed, when applying our technique with the measured strength for the \hb, \mgb, \fei, and \feii\ indices that \citet{McDer15} measure inside one effective radius (11\arcsec) and using the \citet{Tho11} models we find stellar age, metallicity, and alpha-element abundance values that are consistent with what we find in the same region with our own line-strength measurements. In the same region \citet{McDer15} finds older stellar ages when using the \citet{Schi07} models, suggesting that the choice of stellar-population models lies at the heart of this discrepancy.

\section{Discussion and conclusions}
\label{sec:Dis}

The true nature of the kinematically distinct components, such as the ones observed in our two galaxies, is best studied using IFU spectroscopy, combining both photometric and kinematic information. Nevertheless, it is intrinsically difficult to reconstruct the formation mechanism for the retrograde stellar populations, observed in these two early-type galaxies. 

\subsection{NGC~448}
Counter-rotating kinematic components can generally originate from processes that are either external or internal to the host galaxy.   
The dominant formation channel for the assembly of stellar counter-rotating kinematic components is most often attributed to the accretion or reprocessing of gas and subsequent in situ star-formation \citep{Cor14}. 

On the one hand, fresh gas could be captured through a gas rich minor merger. External gas can be acquired from the companion in a retrograde fashion and is of considerate quantity (i.e. more than what is already present within the more massive galaxy) it does not get completely dissipated and swept up by pre-existing gas within the more massive host galaxy and a subsequent star-formation episode results in a population of retrograde stars. This was recently illustrated through the aid of extensive merger numerical simulations by \citet{Bass17} and previously addressed by \citet{Thak96, Tha98}. They demonstrate how similar gas-rich minor mergers are likely to result in an S0 galaxy or did not change the morphological properties of an S0 progenitor. Usually, the resultant post-merger products of such simulations contain some appreciable gas reservoir. However, overall NGC~448 is surprisingly devoid of gas, contrary to other observed galaxies harbouring counter-rotating stellar disks. We confirm the absence of ionised-gas emission in our IFU spectra, as previously reported by \citet{Kat16} in their long-slit spectrum. This indicates that there is no ongoing substantial residual star formation. With these considerations, we cannot unambiguously use the coexistence of a gas reservoir and a counter-rotating disk as evidence that indeed gas accretion and subsequent in situ star-formation formed the truly kinematically decoupled structure. One possible way to reconcile such a formation scenario with our observations is to postulate that the gas has been fully consumed and processed to stars. As a consequence, the resulting recent star-formation episode should have left a sub-population of young stars predominately belonging to the counter-rotating disk. However, we do not detect any significant difference in the ages of the stellar populations of either the main galaxy body or the decoupled kinematic subcomponent. In this case, the flat age radial profile and very sharp negative metallicity gradient could be the result of a rapid outside-in formation. Still, using the recovered alpha-element abundance and eq.(4) from \citet{Tho05} implies that the counter-rotating disk stars formed through an extended period of star formation that lasted at least 10 Gyr. Therefore, the observed low $\alpha$/Fe values in combination with the high metallicity and similar to the main galactic stellar body age of the counter-rotating component are difficult to reconcile in a purely closed box system. The low $\alpha$/Fe values could instead be present because the accreted  gas material had already been iron enriched to lower down the $\alpha$/Fe ratio. Alternatively, star-formation could have proceeded in both of the kinematic components (disk and main stellar body) generating a noticeably large ($\geqslant$ 1\% in mass) and young (t $\leqslant$ 2.5 Gyr) stellar sub-populations. This in turn would impact our line-strength luminosity-weighted age estimates and metallicities, significantly shifting the age of the main stellar body towards smaller age estimates and the metallicity of the kinematically decoupled disk towards the one of its oldest stellar sub-population \citep[e.g.,][]{Ser07, San14}. A more thorough analysis of the stellar population properties allowing for multiple simple stellar populations instead of just one as in this work using a higher resolution spectra could suffice to either confirm or rule out such a speculation. As an alternative to a gas-rich merger, \citet{Thak96, Tha98} studied the possibility that the gas can fall in from the environment surrounding the galaxy through either short or prolonged period of accretion. These numerical simulations produced counter-rotating disks, albeit most often resulting in other than nearly exponential profiles, such as the one we recover for the counter-rotating component in NGC~448. Another possibility is that the galaxy was originally formed through the filamentary gas accretion from two distinct cosmological filaments, in turn forming the two distinct kinematic structures under a special spacial configuration. \citet{Alg14} studied this process in a cosmological simulation of the formation of a disk galaxy. They conclude that a natural consequence of such a formation scenario would be a discriminable age difference in the two distinct kinematic components. As previously pointed out this is not found to be the case for NGC~448. 

A further puzzle results from the likely interaction with a suspected companion galaxy (LEDA 212690). A tidal stream was observed by \citet{Duc15} and also pointed out by \citet{Kat16}. They remarked that the galaxy is in tidal interaction with a morphologically disturbed companion. From the colour difference available as part of the \citet{Duc15} survey, we can only speculate that the age of the stars constituting the tidal tail is similar to that of the stellar population of NGC~448. If the two galaxies are interacting, then the redshift of the suspected companion LEDA 212690 was possibly wrongfully inferred as $\sim 0.074$ as part of the CAIRNS survey \citep{Rin03}. Another more exotic possibility is that the counter-rotating disk was assembled through the direct accretion of stars from a companion. It is however extremely unlikely that such a formation scenario would reproduce the observed metalicity gradient for the counter-rotating disk. There is no strong physical reason for the stars to be accreted preferentially in a spatial configuration that would produce a counter-rotating disk with higher metallicity than the main galaxy body one and with a strong gradient. Further observations of LEDA 212690 would unequivocally rule out some of the previously mentioned external formation scenarios for the counter-rotating disk.

As opposed to the external acquisition of gas and a subsequent star formation in situ, the formation of counter-rotating embedded stellar disk could be the product of an internal formation mechanism. The internal separatrix formation scenario \citep{Eva94} would be very unlikely in the case of NGC~448. Even though the two components have very similar stellar population ages, the counter-rotating disk is less extended than the main galaxy body, contrary to the predictions of such a formation mechanism. Furthermore, the chemical properties of the two stellar populations are quite different in opposition whereas the stars forming the counter-rotating disk should share the same star-formation history as the galaxy. \citet{Kanth16} argued that counter-rotation in general could also stem from the exchange of angular momentum in close galaxy encounters. As pointed out previously we observe a ``bridge'' between the two galaxies. Nevertheless, we do not see any drastic misalignment in the kinematic rotation centres of the counter-rotating embedded disk and the main galactic body and we report a difference in the chemical properties of the stellar populations of the two kinematic components. Also, the two counter-rotating structures seem to share the same axis of rotation according to the accuracy of our velocity measurements (i.e. we do not observe an obvious offset between the heliocentric velocities of the two kinematic components). As the aforementioned angular momentum exchange due to a close fly-by scenario would largely result in the two components sharing different rotation axes we conclude that it is unlikely to be the origin of the separate stellar population kinematic components observed in NGC~448. 

\subsection{NGC~4365}

Even though we retrieved a second photometric component through our photometric decomposition (see, \S~\ref{sec:PhDec}) and also obtained a good spectral decomposition in some of our Voronoi bins (see, \S~\ref{subsec:KD4365}), we did not get a satisfactory consistence in our photometric and kinematic decomposition to support the presence of two truly decoupled kinematic structures. It is well established by means of simulations that the KDCs in massive early-type ellipticals could be a product of a galaxy merger. To complement the finding of the ATLAS$^{\rm 3D}$, \citet{Bo11} produced an extensive set of such merger simulations spanning a long range of possible mass ratios (1:1 to 6:1), initial conditions, and orbital parameters in order to discern the origins of the two main classes of early-type galaxies on the basis of their resultant angular momentum. They report that the majority of the slow-rotator merger products in their simulations possess a distinct KDC. Additionally, a major merger of two disk galaxies even when initially both following prograde orbit \citep{Tsat15} could also produce a KDC in the resultant elliptical galaxy merger product. However, both these KDC formation channels would result in a noticeable age difference compared to the main stellar body. As previously found by \citet{Dav01} and reiterated by \citet{vdBos08}, the ages and properties of both the KDC and the rest of the galaxy have the same magnesium-to-iron abundance ratios and virtually the same ages implying that the KDC should have formed at least 12 Gyr ago or is indeed a projection effect. Furthermore, NGC 4365 does not even show symmetric velocity dispersion peaks and therefore does not resemble a typical $2\sigma$ galaxies, considered to host truly decoupled counter-rotating structures. Even in the presence of such features for NGC 5813 \citet{Kraj15} did not infer that the KDC arises from the counter-rotating of two disks, but through detailed modelling as previously also done in the context of NGC~4365 \citep{vdBos08} found that the KDC is likely not due to a truly structurally or dynamically decoupled components. It was found to most plausibly be the result of the complex nature of the orbits in such quite massive ellipticals.

\subsection{Summary}
We investigated two early-type galaxies NGC~448 and NGC~4365, that show central kinematically distinct cores. Through the kinematic decomposition technique by \citet{Cocc11} we separated the individual contributions of two distinct kinematic components with the aid of integral-field observations of NGC~448. 
NGC~448 hosts a counter-rotating stellar kinematic structure with an angular size of at least 42\arcsec, which is considerably more extended ($\sim$1.5 times) than the regions with irregular stellar kinematics.
Based on the observed nearly exponential photometric profile and the kinematic information we extracted, the distinct component is very kinematically cold ($V_{\star}/\sigma_{\star}$ >> 1), therefore likely having a disk morphology. The disk counter-rotates with similar velocity to that of the main stellar body. The stars of counter-rotating disk stars and main body have comparable age. However, the two distinct kinematic components have chemically different stellar populations. The stellar population of the counter-rotating disk displays steeper metallicity gradient and is slightly less alpha-element enhanced than the one constituting the bulk of the galaxy. In contrast to all other galaxies with embedded counter-rotating disks we do not observe much interstellar gas. Although this prohibits us to, therefore, unequivocally link the formation of the counter-rotating disk to a past gas accretion event or a ``wet'' merger, our analysis points to a gas-rich merger after which the gas reservoir was exhausted in a period of outside-in star-formation episode, provided it was pre-iron-enriched and in retrograde orbits, or simultaneous star-formation bursts in both kinematic components, as the two likely formation mechanisms for the counter-rotating stellar disk. We find no clear signature in the integral-field observations of NGC~4365 for a true kinematic decoupling. The KDC in this galaxy likely stems as previously suggested from a projection effect due to its triaxial nature.  

\begin{acknowledgements}
Both BN and MS acknowledge the financial support and hospitality provided by the European Southern Observatory (ESO) through the award of an ESO studentship and visiting fellowship, respectively. EMC, EDB, LM, and AP acknowledge financial support from Padua University through grants DOR1699945/16, DOR1715817/17, DOR1885254/18, and BIRD164402/16.  
\end{acknowledgements}

\bibliographystyle{aa} 

\begin{thebibliography}{}
\expandafter\ifx\csname natexlab\endcsname\relax\def\natexlab#1{#1}\fi

\bibitem[{{Algorry} {et~al.}(2014){Algorry}, {Navarro}, {Abadi}, {Sales},
  {Steinmetz}, \& {Piontek}}]{Alg14}
{Algorry}, D.~G., {Navarro}, J.~F., {Abadi}, M.~G., {et~al.} 2014, \mnras, 437,
  3596

\bibitem[{{Arnold} {et~al.}(2014){Arnold}, {Romanowsky}, {Brodie}, {Forbes},
  {Strader}, {Spitler}, {Foster}, {Blom}, {Kartha}, {Pastorello}, {Pota},
  {Usher}, \& {Woodley}}]{Arn14}
{Arnold}, J.~A., {Romanowsky}, A.~J., {Brodie}, J.~P., {et~al.} 2014, \apj,
  791, 80

\bibitem[{{Balcells} \& {Quinn}(1990)}]{Balc90}
{Balcells}, M. \& {Quinn}, P.~J. 1990, \apj, 361, 381

\bibitem[{{Bassett} {et~al.}(2017){Bassett}, {Bekki}, {Cortese}, \&
  {Couch}}]{Bass17}
{Bassett}, R., {Bekki}, K., {Cortese}, L., \& {Couch}, W. 2017, \mnras, 471,
  1892

\bibitem[{{Beers} {et~al.}(1990){Beers}, {Flynn}, \& {Gebhardt}}]{Beer90}
{Beers}, T.~C., {Flynn}, K., \& {Gebhardt}, K. 1990, \aj, 100, 32

\bibitem[{{Bender}(1988)}]{Ben88}
{Bender}, R. 1988, \aap, 202, L5

\bibitem[{{Bender}(1990)}]{Ben90}
{Bender}, R. 1990, \aap, 229, 441

\bibitem[{{Bertola} {et~al.}(1996){Bertola}, {Cinzano}, {Corsini}, {Pizzella},
  {Persic}, \& {Salucci}}]{Ber96}
{Bertola}, F., {Cinzano}, P., {Corsini}, E.~M., {et~al.} 1996, \apjl, 458, L67

\bibitem[{{Bois} {et~al.}(2011){Bois}, {Emsellem}, {Bournaud}, {Alatalo},
  {Blitz}, {Bureau}, {Cappellari}, {Davies}, {Davis}, {de Zeeuw}, {Duc},
  {Khochfar}, {Krajnovi{\'c}}, {Kuntschner}, {Lablanche}, {McDermid},
  {Morganti}, {Naab}, {Oosterloo}, {Sarzi}, {Scott}, {Serra}, {Weijmans}, \&
  {Young}}]{Bo11}
{Bois}, M., {Emsellem}, E., {Bournaud}, F., {et~al.} 2011, \mnras, 416, 1654

\bibitem[{{Cappellari}(2016)}]{Cap16}
{Cappellari}, M. 2016, \araa, 54, 597

\bibitem[{{Cappellari}(2017)}]{Cap17}
{Cappellari}, M. 2017, \mnras, 466, 798

\bibitem[{{Cappellari} \& {Copin}(2003)}]{Cap03}
{Cappellari}, M. \& {Copin}, Y. 2003, \mnras, 342, 345

\bibitem[{{Cappellari} \& {Emsellem}(2004)}]{Cap04}
{Cappellari}, M. \& {Emsellem}, E. 2004, \pasp, 116, 138

\bibitem[{{Cappellari} {et~al.}(2007){Cappellari}, {Emsellem}, {Bacon},
  {Bureau}, {Davies}, {de Zeeuw}, {Falc{\'o}n-Barroso}, {Krajnovi{\'c}},
  {Kuntschner}, {McDermid}, {Peletier}, {Sarzi}, {van den Bosch}, \& {van de
  Ven}}]{Cap07}
{Cappellari}, M., {Emsellem}, E., {Bacon}, R., {et~al.} 2007, \mnras, 379, 418

\bibitem[{{Cappellari} {et~al.}(2011){Cappellari}, {Emsellem}, {Krajnovi{\'c}},
  {McDermid}, {Scott}, {Verdoes Kleijn}, {Young}, {Alatalo}, {Bacon}, {Blitz},
  {Bois}, {Bournaud}, {Bureau}, {Davies}, {Davis}, {de Zeeuw}, {Duc},
  {Khochfar}, {Kuntschner}, {Lablanche}, {Morganti}, {Naab}, {Oosterloo},
  {Sarzi}, {Serra}, \& {Weijmans}}]{Cap11}
{Cappellari}, M., {Emsellem}, E., {Krajnovi{\'c}}, D., {et~al.} 2011, \mnras,
  413, 813

\bibitem[{{Cardiel} {et~al.}(1998){Cardiel}, {Gorgas}, {Cenarro}, \&
  {Gonzalez}}]{Car98}
{Cardiel}, N., {Gorgas}, J., {Cenarro}, J., \& {Gonzalez}, J.~J. 1998, \aaps,
  127, 597

\bibitem[{{Chilingarian} {et~al.}(2007{\natexlab{a}}){Chilingarian},
  {Prugniel}, {Sil'Chenko}, \& {Koleva}}]{Chil07a}
{Chilingarian}, I., {Prugniel}, P., {Sil'Chenko}, O., \& {Koleva}, M.
  2007{\natexlab{a}}, in IAU Symposium, Vol. 241, Stellar Populations as
  Building Blocks of Galaxies, ed. A.~{Vazdekis} \& R.~{Peletier}, 175--176

\bibitem[{{Chilingarian} {et~al.}(2007{\natexlab{b}}){Chilingarian},
  {Prugniel}, {Sil'Chenko}, \& {Afanasiev}}]{Chil07}
{Chilingarian}, I.~V., {Prugniel}, P., {Sil'Chenko}, O.~K., \& {Afanasiev},
  V.~L. 2007{\natexlab{b}}, \mnras, 376, 1033

\bibitem[{{Coccato} {et~al.}(2015){Coccato}, {Fabricius}, {Morelli}, {Corsini},
  {Pizzella}, {Erwin}, {Dalla Bont{\`a}}, {Saglia}, {Bender}, \&
  {Williams}}]{Cocc15}
{Coccato}, L., {Fabricius}, M., {Morelli}, L., {et~al.} 2015, \aap, 581, A65

\bibitem[{{Coccato} {et~al.}(2014){Coccato}, {Iodice}, \& {Arnaboldi}}]{Cocc14}
{Coccato}, L., {Iodice}, E., \& {Arnaboldi}, M. 2014, \aap, 569, A83

\bibitem[{{Coccato} {et~al.}(2011){Coccato}, {Morelli}, {Corsini}, {Buson},
  {Pizzella}, {Vergani}, \& {Bertola}}]{Cocc11}
{Coccato}, L., {Morelli}, L., {Corsini}, E.~M., {et~al.} 2011, \mnras, 412,
  L113

\bibitem[{{Coccato} {et~al.}(2013){Coccato}, {Morelli}, {Pizzella}, {Corsini},
  {Buson}, \& {Dalla Bont{\`a}}}]{Cocc13}
{Coccato}, L., {Morelli}, L., {Pizzella}, A., {et~al.} 2013, \aap, 549, A3

\bibitem[{{Corsini}(2014)}]{Cor14}
{Corsini}, E.~M. 2014, in Astronomical Society of the Pacific Conference
  Series, Vol. 486, Multi-Spin Galaxies, ed. E.~{Iodice} \& E.~M. {Corsini}, 51

\bibitem[{{Crocker} {et~al.}(2009){Crocker}, {Jeong}, {Komugi}, {Combes},
  {Bureau}, {Young}, \& {Yi}}]{Cro09}
{Crocker}, A.~F., {Jeong}, H., {Komugi}, S., {et~al.} 2009, \mnras, 393, 1255

\bibitem[{{Davies} {et~al.}(2001){Davies}, {Kuntschner}, {Emsellem}, {Bacon},
  {Bureau}, {Carollo}, {Copin}, {Miller}, {Monnet}, {Peletier}, {Verolme}, \&
  {de Zeeuw}}]{Dav01}
{Davies}, R.~L., {Kuntschner}, H., {Emsellem}, E., {et~al.} 2001, \apjl, 548,
  L33

\bibitem[{{de Zeeuw} \& {Franx}(1991)}]{deZ91}
{de Zeeuw}, T. \& {Franx}, M. 1991, \araa, 29, 239

\bibitem[{{Duc} {et~al.}(2015){Duc}, {Cuillandre}, {Karabal}, {Cappellari},
  {Alatalo}, {Blitz}, {Bournaud}, {Bureau}, {Crocker}, {Davies}, {Davis}, {de
  Zeeuw}, {Emsellem}, {Khochfar}, {Krajnovi{\'c}}, {Kuntschner}, {McDermid},
  {Michel-Dansac}, {Morganti}, {Naab}, {Oosterloo}, {Paudel}, {Sarzi}, {Scott},
  {Serra}, {Weijmans}, \& {Young}}]{Duc15}
{Duc}, P.-A., {Cuillandre}, J.-C., {Karabal}, E., {et~al.} 2015, \mnras, 446,
  120

\bibitem[{{Evans} \& {Collett}(1994)}]{Eva94}
{Evans}, N.~W. \& {Collett}, J.~L. 1994, \apjl, 420, L67

\bibitem[{{Ferrarese} {et~al.}(2006){Ferrarese}, {C{\^o}t{\'e}}, {Jord{\'a}n},
  {Peng}, {Blakeslee}, {Piatek}, {Mei}, {Merritt}, {Milosavljevi{\'c}},
  {Tonry}, \& {West}}]{Fer06}
{Ferrarese}, L., {C{\^o}t{\'e}}, P., {Jord{\'a}n}, A., {et~al.} 2006, \apjs,
  164, 334

\bibitem[{{Franx} \& {Illingworth}(1988)}]{Fra88}
{Franx}, M. \& {Illingworth}, G.~D. 1988, \apjl, 327, L55

\bibitem[{{Freudling} {et~al.}(2013){Freudling}, {Romaniello}, {Bramich},
  {Ballester}, {Forchi}, {Garc{\'{\i}}a-Dabl{\'o}}, {Moehler}, \&
  {Neeser}}]{Freud13}
{Freudling}, W., {Romaniello}, M., {Bramich}, D.~M., {et~al.} 2013, \aap, 559,
  A96

\bibitem[{{Girardi} {et~al.}(2000){Girardi}, {Bressan}, {Bertelli}, \&
  {Chiosi}}]{Gir00}
{Girardi}, L., {Bressan}, A., {Bertelli}, G., \& {Chiosi}, C. 2000, \aaps, 141,
  371

\bibitem[{{Gorgas} {et~al.}(1990){Gorgas}, {Efstathiou}, \& {Aragon
  Salamanca}}]{Gorg90}
{Gorgas}, J., {Efstathiou}, G., \& {Aragon Salamanca}, A. 1990, \mnras, 245,
  217

\bibitem[{{Gu{\'e}rou} {et~al.}(2016){Gu{\'e}rou}, {Emsellem}, {Krajnovi{\'c}},
  {McDermid}, {Contini}, \& {Weilbacher}}]{Gue16}
{Gu{\'e}rou}, A., {Emsellem}, E., {Krajnovi{\'c}}, D., {et~al.} 2016, \aap,
  591, A143

\bibitem[{{Hoffman} {et~al.}(2010){Hoffman}, {Cox}, {Dutta}, \&
  {Hernquist}}]{Hoff10}
{Hoffman}, L., {Cox}, T.~J., {Dutta}, S., \& {Hernquist}, L. 2010, \apj, 723,
  818

\bibitem[{{Iodice} {et~al.}(2015){Iodice}, {Coccato}, {Combes}, {de Zeeuw},
  {Arnaboldi}, {Weilbacher}, {Bacon}, {Kuntschner}, \& {Spavone}}]{Iod15}
{Iodice}, E., {Coccato}, L., {Combes}, F., {et~al.} 2015, \aap, 583, A48

\bibitem[{{Jedrzejewski}(1987)}]{Jed87}
{Jedrzejewski}, R.~I. 1987, \mnras, 226, 747

\bibitem[{{Johnston} {et~al.}(2013){Johnston}, {Merrifield},
  {Arag{\'o}n-Salamanca}, \& {Cappellari}}]{John13}
{Johnston}, E.~J., {Merrifield}, M.~R., {Arag{\'o}n-Salamanca}, A., \&
  {Cappellari}, M. 2013, \mnras, 428, 1296

\bibitem[{{Kantharia}(2016)}]{Kanth16}
{Kantharia}, N.~G. 2016, ArXiv e-prints [\eprint[arXiv]{1606.04242}]

\bibitem[{{Katkov} {et~al.}(2013){Katkov}, {Sil'chenko}, \&
  {Afanasiev}}]{Kat13}
{Katkov}, I.~Y., {Sil'chenko}, O.~K., \& {Afanasiev}, V.~L. 2013, \apj, 769,
  105

\bibitem[{{Katkov} {et~al.}(2016){Katkov}, {Sil'chenko}, {Chilingarian},
  {Uklein}, \& {Egorov}}]{Kat16}
{Katkov}, I.~Y., {Sil'chenko}, O.~K., {Chilingarian}, I.~V., {Uklein}, R.~I.,
  \& {Egorov}, O.~V. 2016, \mnras, 461, 2068

\bibitem[{{Kormendy}(1984)}]{Kor84}
{Kormendy}, J. 1984, \apj, 287, 577

\bibitem[{{Krajnovi{\'c}} {et~al.}(2013){Krajnovi{\'c}}, {Alatalo}, {Blitz},
  {Bois}, {Bournaud}, {Bureau}, {Cappellari}, {Davies}, {Davis}, {de Zeeuw},
  {Duc}, {Emsellem}, {Khochfar}, {Kuntschner}, {McDermid}, {Morganti}, {Naab},
  {Oosterloo}, {Sarzi}, {Scott}, {Serra}, {Weijmans}, \& {Young}}]{Kraj13}
{Krajnovi{\'c}}, D., {Alatalo}, K., {Blitz}, L., {et~al.} 2013, \mnras, 432,
  1768

\bibitem[{{Krajnovi{\'c}} {et~al.}(2008){Krajnovi{\'c}}, {Bacon}, {Cappellari},
  {Davies}, {de Zeeuw}, {Emsellem}, {Falc{\'o}n-Barroso}, {Kuntschner},
  {McDermid}, {Peletier}, {Sarzi}, {van den Bosch}, \& {van de Ven}}]{Kraj08}
{Krajnovi{\'c}}, D., {Bacon}, R., {Cappellari}, M., {et~al.} 2008, \mnras, 390,
  93

\bibitem[{{Krajnovi{\'c}} {et~al.}(2015){Krajnovi{\'c}}, {Weilbacher},
  {Urrutia}, {Emsellem}, {Carollo}, {Shirazi}, {Bacon}, {Contini}, {Epinat},
  {Kamann}, {Martinsson}, \& {Steinmetz}}]{Kraj15}
{Krajnovi{\'c}}, D., {Weilbacher}, P.~M., {Urrutia}, T., {et~al.} 2015, \mnras,
  452, 2

\bibitem[{Krajnović {et~al.}(2011)Krajnović, Emsellem, Cappellari, Alatalo,
  Blitz, Bois, Bournaud, Bureau, Davies, Davis, de~Zeeuw, Khochfar, Kuntschner,
  Lablanche, McDermid, Morganti, Naab, Oosterloo, Sarzi, Scott, Serra,
  Weijmans, \& Young}]{Kraj11}
Krajnović, D., Emsellem, E., Cappellari, M., {et~al.} 2011, \mnras, 414, 2923

\bibitem[{{Le Borgne} {et~al.}(2004){Le Borgne}, {Rocca-Volmerange},
  {Prugniel}, {Lan{\c c}on}, {Fioc}, \& {Soubiran}}]{LeBorg04}
{Le Borgne}, D., {Rocca-Volmerange}, B., {Prugniel}, P., {et~al.} 2004, \aap,
  425, 881

\bibitem[{{Markwardt}(2009)}]{Mar09}
{Markwardt}, C.~B. 2009, in Astronomical Society of the Pacific Conference
  Series, Vol. 411, Astronomical Data Analysis Software and Systems XVIII, ed.
  D.~A. {Bohlender}, D.~{Durand}, \& P.~{Dowler}, 251

\bibitem[{{McDermid} {et~al.}(2015){McDermid}, {Alatalo}, {Blitz}, {Bournaud},
  {Bureau}, {Cappellari}, {Crocker}, {Davies}, {Davis}, {de Zeeuw}, {Duc},
  {Emsellem}, {Khochfar}, {Krajnovi{\'c}}, {Kuntschner}, {Morganti}, {Naab},
  {Oosterloo}, {Sarzi}, {Scott}, {Serra}, {Weijmans}, \& {Young}}]{McDer15}
{McDermid}, R.~M., {Alatalo}, K., {Blitz}, L., {et~al.} 2015, \mnras, 448, 3484

\bibitem[{{McDermid} {et~al.}(2006){McDermid}, {Emsellem}, {Shapiro}, {Bacon},
  {Bureau}, {Cappellari}, {Davies}, {de Zeeuw}, {Falc{\'o}n-Barroso},
  {Krajnovi{\'c}}, {Kuntschner}, {Peletier}, \& {Sarzi}}]{McDer06}
{McDermid}, R.~M., {Emsellem}, E., {Shapiro}, K.~L., {et~al.} 2006, \mnras,
  373, 906

\bibitem[{{Mitzkus} {et~al.}(2017){Mitzkus}, {Cappellari}, \&
  {Walcher}}]{Mitzk17}
{Mitzkus}, M., {Cappellari}, M., \& {Walcher}, C.~J. 2017, \mnras, 464, 4789

\bibitem[{{Morelli} {et~al.}(2012){Morelli}, {Corsini}, {Pizzella}, {Dalla
  Bont{\`a}}, {Coccato}, {M{\'e}ndez-Abreu}, \& {Cesetti}}]{Mor12}
{Morelli}, L., {Corsini}, E.~M., {Pizzella}, A., {et~al.} 2012, \mnras, 423,
  962

\bibitem[{{Morelli} {et~al.}(2017){Morelli}, {Pizzella}, {Coccato}, {Corsini},
  {Dalla Bont{\`a}}, {Buson}, {Ivanov}, {Pagotto}, {Pompei}, \&
  {Rocco}}]{More17}
{Morelli}, L., {Pizzella}, A., {Coccato}, L., {et~al.} 2017, \aap, 600, A76

\bibitem[{Nelder \& Mead(1965)}]{Nel65}
Nelder, J.~A. \& Mead, R. 1965, The Computer Journal, 7, 308

\bibitem[{{Nyland} {et~al.}(2017){Nyland}, {Young}, {Wrobel}, {Davis},
  {Bureau}, {Alatalo}, {Morganti}, {Duc}, {de Zeeuw}, {McDermid}, {Crocker}, \&
  {Oosterloo}}]{Nyl17}
{Nyland}, K., {Young}, L.~M., {Wrobel}, J.~M., {et~al.} 2017, \mnras, 464, 1029

\bibitem[{{Peng} {et~al.}(2002){Peng}, {Ho}, {Impey}, \& {Rix}}]{Peng02}
{Peng}, C.~Y., {Ho}, L.~C., {Impey}, C.~D., \& {Rix}, H.-W. 2002, \aj, 124, 266

\bibitem[{{Pizzella} {et~al.}(2004){Pizzella}, {Corsini}, {Vega Beltr{\'a}n},
  \& {Bertola}}]{Pizz04}
{Pizzella}, A., {Corsini}, E.~M., {Vega Beltr{\'a}n}, J.~C., \& {Bertola}, F.
  2004, \aap, 424, 447

\bibitem[{{Pizzella} {et~al.}(2014){Pizzella}, {Morelli}, {Corsini}, {Dalla
  Bont{\`a}}, {Coccato}, \& {Sanjana}}]{Pizz14}
{Pizzella}, A., {Morelli}, L., {Corsini}, E.~M., {et~al.} 2014, \aap, 570, A79

\bibitem[{{Puerari} \& {Pfenniger}(2001)}]{Pue01}
{Puerari}, I. \& {Pfenniger}, D. 2001, \apss, 276, 909

\bibitem[{{Rines} {et~al.}(2003){Rines}, {Geller}, {Kurtz}, \&
  {Diaferio}}]{Rin03}
{Rines}, K., {Geller}, M.~J., {Kurtz}, M.~J., \& {Diaferio}, A. 2003, \aj, 126,
  2152

\bibitem[{{Rix} {et~al.}(1992){Rix}, {Franx}, {Fisher}, \&
  {Illingworth}}]{Rix92}
{Rix}, H.-W., {Franx}, M., {Fisher}, D., \& {Illingworth}, G. 1992, \apjl, 400,
  L5

\bibitem[{{Rix} \& {White}(1992)}]{Rix92a}
{Rix}, H.-W. \& {White}, S.~D.~M. 1992, \mnras, 254, 389

\bibitem[{{Rubin}(1994)}]{Rub94}
{Rubin}, V.~C. 1994, \aj, 108, 456

\bibitem[{{Rubin} {et~al.}(1992){Rubin}, {Graham}, \& {Kenney}}]{Rub92}
{Rubin}, V.~C., {Graham}, J.~A., \& {Kenney}, J.~D.~P. 1992, \apjl, 394, L9

\bibitem[{{Salpeter}(1955)}]{Sal55}
{Salpeter}, E.~E. 1955, \apj, 121, 161

\bibitem[{{S{\'a}nchez-Bl{\'a}zquez} {et~al.}(2014){S{\'a}nchez-Bl{\'a}zquez},
  {Rosales-Ortega}, {M{\'e}ndez-Abreu}, {P{\'e}rez}, {S{\'a}nchez}, {Zibetti},
  {Aguerri}, {Bland-Hawthorn}, {Catal{\'a}n-Torrecilla}, {Cid Fernandes}, {de
  Amorim}, {de Lorenzo-Caceres}, {Falc{\'o}n-Barroso}, {Galazzi},
  {Garc{\'{\i}}a Benito}, {Gil de Paz}, {Gonz{\'a}lez Delgado}, {Husemann},
  {Iglesias-P{\'a}ramo}, {Jungwiert}, {Marino}, {M{\'a}rquez}, {Mast},
  {Mendoza}, {Moll{\'a}}, {Papaderos}, {Ruiz-Lara}, {van de Ven}, {Walcher}, \&
  {Wisotzki}}]{San14}
{S{\'a}nchez-Bl{\'a}zquez}, P., {Rosales-Ortega}, F.~F., {M{\'e}ndez-Abreu},
  J., {et~al.} 2014, \aap, 570, A6

\bibitem[{{Sarzi} {et~al.}(2006){Sarzi}, {Falc{\'o}n-Barroso}, {Davies},
  {Bacon}, {Bureau}, {Cappellari}, {de Zeeuw}, {Emsellem}, {Fathi},
  {Krajnovi{\'c}}, {Kuntschner}, {McDermid}, \& {Peletier}}]{Sar06}
{Sarzi}, M., {Falc{\'o}n-Barroso}, J., {Davies}, R.~L., {et~al.} 2006, \mnras,
  366, 1151

\bibitem[{{Sarzi} {et~al.}(2018){Sarzi}, {Iodice}, {Coccato}, {Corsini}, {de
  Zeeuw}, {Falc{\'o}n-Barroso}, {Gadotti}, {Lyubenova}, {McDermid}, {van de
  Ven}, {Fahrion}, {Pizzella}, \& {Zhu}}]{Sar18}
{Sarzi}, M., {Iodice}, E., {Coccato}, L., {et~al.} 2018, ArXiv e-prints
  [\eprint[arXiv]{1804.06795}]

\bibitem[{{Sarzi} {et~al.}(2016){Sarzi}, {Ledo}, {Coccato}, {Corsini}, {Dotti},
  {Khochfar}, {Maraston}, {Morelli}, \& {Pizzella}}]{Sar16}
{Sarzi}, M., {Ledo}, H.~R., {Coccato}, L., {et~al.} 2016, \mnras, 457, 1804

\bibitem[{{Schiavon}(2007)}]{Schi07}
{Schiavon}, R.~P. 2007, \apjs, 171, 146

\bibitem[{{Schulze} {et~al.}(2017){Schulze}, {Remus}, \& {Dolag}}]{Sch17}
{Schulze}, F., {Remus}, R.-S., \& {Dolag}, K. 2017, Galaxies, 5, 41

\bibitem[{{Schwarzschild}(1979)}]{Schw79}
{Schwarzschild}, M. 1979, \apj, 232, 236

\bibitem[{{Serra} \& {Trager}(2007)}]{Ser07}
{Serra}, P. \& {Trager}, S.~C. 2007, \mnras, 374, 769

\bibitem[{{Soto} {et~al.}(2016){Soto}, {Lilly}, {Bacon}, {Richard}, \&
  {Conseil}}]{Sot16b}
{Soto}, K.~T., {Lilly}, S.~J., {Bacon}, R., {Richard}, J., \& {Conseil}, S.
  2016, \mnras, 458, 3210

\bibitem[{{Statler}(1991)}]{Sta91}
{Statler}, T.~S. 1991, \aj, 102, 882

\bibitem[{{Tabor} {et~al.}(2017){Tabor}, {Merrifield}, {Arag{\'o}n-Salamanca},
  {Cappellari}, {Bamford}, \& {Johnston}}]{Tab17}
{Tabor}, M., {Merrifield}, M., {Arag{\'o}n-Salamanca}, A., {et~al.} 2017,
  \mnras, 466, 2024

\bibitem[{{Thakar} \& {Ryden}(1996)}]{Thak96}
{Thakar}, A.~R. \& {Ryden}, B.~S. 1996, \apj, 461, 55

\bibitem[{{Thakar} \& {Ryden}(1998)}]{Tha98}
{Thakar}, A.~R. \& {Ryden}, B.~S. 1998, \apj, 506, 93

\bibitem[{{Thomas} {et~al.}(2003){Thomas}, {Maraston}, \& {Bender}}]{Thom03}
{Thomas}, D., {Maraston}, C., \& {Bender}, R. 2003, \mnras, 339, 897

\bibitem[{{Thomas} {et~al.}(2005){Thomas}, {Maraston}, {Bender}, \& {Mendes de
  Oliveira}}]{Tho05}
{Thomas}, D., {Maraston}, C., {Bender}, R., \& {Mendes de Oliveira}, C. 2005,
  \apj, 621, 673

\bibitem[{{Thomas} {et~al.}(2011){Thomas}, {Maraston}, \& {Johansson}}]{Tho11}
{Thomas}, D., {Maraston}, C., \& {Johansson}, J. 2011, \mnras, 412, 2183

\bibitem[{{Tsatsi} {et~al.}(2015){Tsatsi}, {Macci{\`o}}, {van de Ven}, \&
  {Moster}}]{Tsat15}
{Tsatsi}, A., {Macci{\`o}}, A.~V., {van de Ven}, G., \& {Moster}, B.~P. 2015,
  \apjl, 802, L3

\bibitem[{{van den Bosch} {et~al.}(2008){van den Bosch}, {van de Ven},
  {Verolme}, {Cappellari}, \& {de Zeeuw}}]{vdBos08}
{van den Bosch}, R.~C.~E., {van de Ven}, G., {Verolme}, E.~K., {Cappellari},
  M., \& {de Zeeuw}, P.~T. 2008, \mnras, 385, 647

\bibitem[{{van der Marel} \& {Franx}(1993)}]{vdMar93}
{van der Marel}, R.~P. \& {Franx}, M. 1993, \apj, 407, 525

\bibitem[{{Vazdekis} {et~al.}(2012){Vazdekis}, {Ricciardelli}, {Cenarro},
  {Rivero-Gonz{\'a}lez}, {D{\'{\i}}az-Garc{\'{\i}}a}, \&
  {Falc{\'o}n-Barroso}}]{Vazd12}
{Vazdekis}, A., {Ricciardelli}, E., {Cenarro}, A.~J., {et~al.} 2012, \mnras,
  424, 157

\bibitem[{{Vergani} {et~al.}(2007){Vergani}, {Pizzella}, {Corsini}, {van
  Driel}, {Buson}, {Dettmar}, \& {Bertola}}]{Verg07}
{Vergani}, D., {Pizzella}, A., {Corsini}, E.~M., {et~al.} 2007, \aap, 463, 883

\bibitem[{{Weilbacher} {et~al.}(2012){Weilbacher}, {Streicher}, {Urrutia},
  {Jarno}, {P{\'e}contal-Rousset}, {Bacon}, \& {B{\"o}hm}}]{Weil12}
{Weilbacher}, P.~M., {Streicher}, O., {Urrutia}, T., {et~al.} 2012, in
  \procspie, Vol. 8451, Software and Cyberinfrastructure for Astronomy II,
  84510B

\bibitem[{{Worthey} {et~al.}(1994){Worthey}, {Faber}, {Gonzalez}, \&
  {Burstein}}]{Wor94}
{Worthey}, G., {Faber}, S.~M., {Gonzalez}, J.~J., \& {Burstein}, D. 1994,
  \apjs, 94, 687

\bibitem[{{Young} {et~al.}(2011){Young}, {Bureau}, {Davis}, {Combes},
  {McDermid}, {Alatalo}, {Blitz}, {Bois}, {Bournaud}, {Cappellari}, {Davies},
  {de Zeeuw}, {Emsellem}, {Khochfar}, {Krajnovi{\'c}}, {Kuntschner},
  {Lablanche}, {Morganti}, {Naab}, {Oosterloo}, {Sarzi}, {Scott}, {Serra}, \&
  {Weijmans}}]{Young11}
{Young}, L.~M., {Bureau}, M., {Davis}, T.~A., {et~al.} 2011, \mnras, 414, 940

\end{thebibliography}
\end{document}